%% file: main.tex
\documentclass[%
 reprint,
superscriptaddress,
 amsmath,amssymb,
 aps,
prstab,
]{revtex4-2}

\usepackage{graphicx}
\usepackage{dcolumn}
\usepackage{bm}
\usepackage{hyperref}
\hypersetup{
    colorlinks=true,
    linkcolor=blue,
    urlcolor=blue,
    citecolor=blue}

\usepackage{siunitx}
\usepackage{xcolor}

\begin{document}

\preprint{APS/123-QED}

\title{Wakefield Generation in Hydrogen and Lithium Plasmas at FACET-II: Diagnostics and First Beam-Plasma Interaction Results}

\input{authorlist}

\date{\today}

\begin{abstract}
    \input{abstract}
\end{abstract}

\maketitle

\input{body.tex}

\nocite{*}
\bibliography{references}

\end{document}

%% file: authorlist.tex

\author{D. Storey}
 \email{dstorey@slac.stanford.edu}
\affiliation{SLAC National Accelerator Laboratory, Menlo Park, CA 94025}%

\author{C. Zhang}
\affiliation{University of California Los Angeles, Los Angeles, CA 90095}%

\author{P. San Miguel Claveria}
\affiliation{LOA, ENSTA Paris, CNRS, École Polytechnique, Institut Polytechnique de Paris, 91762 Palaiseau, France}

\author{G. J. Cao}
\affiliation{University of Oslo, 0316 Oslo, Norway}

\author{E. Adli}
\affiliation{University of Oslo, 0316 Oslo, Norway}

\author{L. Alsberg}
\affiliation{SLAC National Accelerator Laboratory, Menlo Park, CA 94025}

\author{R. Ariniello}
\affiliation{SLAC National Accelerator Laboratory, Menlo Park, CA 94025}

\author{C. Clarke}
\affiliation{SLAC National Accelerator Laboratory, Menlo Park, CA 94025}

\author{S. Corde}
\affiliation{LOA, ENSTA Paris, CNRS, École Polytechnique, Institut Polytechnique de Paris, 91762 Palaiseau, France}

\author{T. N. Dalichaouch}
\affiliation{University of California Los Angeles, Los Angeles, CA 90095}%

\author{C. E. Doss}
\affiliation{University of Colorado Boulder, Boulder, Colorado 80309}

\author{H. Ekerfelt}
\affiliation{SLAC National Accelerator Laboratory, Menlo Park, CA 94025}

\author{C. Emma}
\affiliation{SLAC National Accelerator Laboratory, Menlo Park, CA 94025}

\author{E. Gerstmayr}
\affiliation{SLAC National Accelerator Laboratory, Menlo Park, CA 94025}
\affiliation{Stanford Pulse Institute, Menlo Park, CA 94305}

\author{S. Gessner}
\affiliation{SLAC National Accelerator Laboratory, Menlo Park, CA 94025}

\author{M. Gilljohann}
\affiliation{LOA, ENSTA Paris, CNRS, École Polytechnique, Institut Polytechnique de Paris, 91762 Palaiseau, France}

\author{C. Hast}
\affiliation{SLAC National Accelerator Laboratory, Menlo Park, CA 94025}

\author{A. Knetsch}
\affiliation{SLAC National Accelerator Laboratory, Menlo Park, CA 94025}
\affiliation{LOA, ENSTA Paris, CNRS, École Polytechnique, Institut Polytechnique de Paris, 91762 Palaiseau, France}

\author{V. Lee}
\affiliation{University of Colorado Boulder, Boulder, Colorado 80309}

\author{M. Litos}
\affiliation{University of Colorado Boulder, Boulder, Colorado 80309}

\author{R. Loney}
\affiliation{SLAC National Accelerator Laboratory, Menlo Park, CA 94025}

\author{K. A. Marsh}
\affiliation{University of California Los Angeles, Los Angeles, CA 90095}%

\author{A. Matheron}
\affiliation{LOA, ENSTA Paris, CNRS, École Polytechnique, Institut Polytechnique de Paris, 91762 Palaiseau, France}

\author{W. B.  Mori}
\affiliation{University of California Los Angeles, Los Angeles, CA 90095}%

\author{Z. Nie}
\affiliation{University of California Los Angeles, Los Angeles, CA 90095}%

\author{B. O'Shea}
\affiliation{SLAC National Accelerator Laboratory, Menlo Park, CA 94025}

\author{M. Parker}
\affiliation{SLAC National Accelerator Laboratory, Menlo Park, CA 94025}

\author{G. White}
\affiliation{SLAC National Accelerator Laboratory, Menlo Park, CA 94025}

\author{G. Yocky}
\affiliation{SLAC National Accelerator Laboratory, Menlo Park, CA 94025}

\author{V. Zakharova}
\affiliation{LOA, ENSTA Paris, CNRS, École Polytechnique, Institut Polytechnique de Paris, 91762 Palaiseau, France}

\author{M. J. Hogan}%
\affiliation{SLAC National Accelerator Laboratory, Menlo Park, CA 94025}

\author{C. Joshi}
\affiliation{University of California Los Angeles, Los Angeles, CA 90095}%

%% file: abstract.tex
Plasma Wakefield Acceleration (PWFA) provides ultrahigh acceleration gradients of 10s of GeV/m, providing a novel path towards efficient, compact, TeV-scale linear colliders and high brightness free electron lasers. 
Critical to the success of these applications is demonstrating simultaneously high gradient acceleration, high energy transfer efficiency, and preservation of emittance, charge, and energy spread.
Experiments at the FACET-II National User Facility at SLAC National Accelerator Laboratory aim to achieve all of these milestones in a single stage plasma wakefield accelerator, providing a \SI{10}{\giga\electronvolt} energy gain in a $<\SI{1}{\meter}$ plasma with high energy transfer efficiency.
Such a demonstration depends critically on diagnostics able to measure emittance with mm-mrad accuracy, energy spectra to determine both \%-level energy spread and broadband energy gain and loss, incoming longitudinal phase space, and matching dynamics.
This paper discusses the experimental setup at FACET-II, including the incoming beam parameters from the FACET-II linac, plasma sources, and diagnostics developed to meet this challenge.
Initial progress on the generation of beam ionized wakes in meter-scale hydrogen gas is discussed, as well as commissioning of the plasma sources and diagnostics.

%% file: body.tex
\section{Introduction}

The pursuit of higher energy and brightness particle beams in the high energy physics and light source communities has pushed conventional accelerator technology to its physical limits. Plans for the next generation TeV-scale linear collider using RF acceleration require extensive lengths of 10s of km~\cite{ILC_TDR, CCC, CLIC} to reach an energy of several \SI{100}{GeV} required for a Higgs factory, and many 10s of km to reach energies greater than \SI{1}{TeV} for energy frontier studies~\cite{SMstudies}.
Plasma Wakefield Acceleration (PWFA) offers a more compact alternative by providing acceleration gradients that are orders of magnitude greater than conventional RF accelerators, opening the door to smaller and more efficient TeV scale electron-positron colliders and free electron lasers.

In a plasma wakefield accelerator, a trailing particle bunch is accelerated by the \textit{wake} left behind a driving relativistic particle beam~\cite{Chen_PRL85} or laser pulse~\cite{Tajima_PRL79} as they propagate through a plasma together.
The driver's transverse fields expel the plasma electrons away from the axis of motion, forming a wake within the previously uniform plasma. These expelled electrons are attracted back towards the axis by the relatively stationary plasma ions, creating a bubble devoid of electrons within the plasma with dimensions on the order of a plasma wavelength, $\lambda_p(\mbox{cm}) = 3.3\times10^6n_p^{-1/2}$, where $n_p$ is the plasma density in \SI{}{cm^{-3}}.
The plasma acts as a transformer, extracting energy from the driver through the formation of a wake, and transferring the wake energy to a trailing bunch. 

To be viable for applications such as linear colliders and high-brightness light sources, plasma wakefield accelerators must deliver high-energy bunches with low energy spread and emittance at high efficiency and repetition rate.
Beam-driven PWFA has achieved significant milestones towards these goals in recent years, including the demonstration of multi-GeV/m accelerating gradients~\cite{Blumenfeld_Nature07, Litos_PPCF2016}, efficient acceleration of narrow energy spread beams~\cite{Lindstrom_PRL21}, and emittance preservation~\cite{Lindstrom_Emit22}.  However, simultaneously achieving all of these parameters--high energy gain, high efficiency, low energy spread, and low emittance--in a plasma wakefield accelerator has yet to be demonstrated.

The upgraded FACET-II National User Facility~\cite{Yakimenko_PRAB19} provides the opportunity for the development of a single stage plasma wakefield accelerator that approaches the parameters required by a future linear collider~\cite{Joshi_2018}.
We aim to demonstrate simultaneously all of the following in a single PWFA stage: energy depletion of the drive bunch energy with a drive-to-wake efficiency of $>80\%$, acceleration of the trailing bunch by $>\SI{10}{\giga\electronvolt}$ with a wake-to-trailing bunch energy extraction efficiency of over 40\%, while simultaneously maintaining beam quality by achieving a final energy spread of $<2\%$ and emittance preservation after acceleration.

This paper describes the beam-driven PWFA approach and the experimental setup at the FACET-II National User Facility which will be used in the demonstration of a single stage plasma accelerator. We also introduce the diagnostics required to demonstrate preserved beam quality, and discuss the initial results obtained from beam-ionized hydrogen plasma studies during user-assisted commissioning of the facility.

\section{Two bunch beam-driven plasma wakefield acceleration}


For the successful application of PWFA in a linear collider, it is crucial to achieve a high energy transfer efficiency from drive to trailing bunch to maximise the overall wall-plug efficiency.
The energy transfer efficiency in PWFA can be considered in two parts -- the \textit{drive-to-wake} efficiency, and \textit{wake-to-trailing bunch} efficiency.
The drive-to-wake efficiency is influenced by several factors, including ensuring sufficient plasma length for near complete energy depletion of the driver, proper matching conditions into the plasma, and the re-acceleration of energy-depleted drive electrons that may slip back into the accelerating phase. Recent experiments at FLASHForward have demonstrated up to 56\% drive-to-wake efficiency~\cite{Pena_23}.
The wake-to-trailing bunch efficiency can be optimized by appropriate beam-loading of the plasma wake and matching into the plasma, with recent experiments demonstrating up to 42\% drive to trailing bunch energy transfer efficiency with preservation of energy spread~\cite{Lindstrom_PRL21}.

The preservation of overall beam quality in PWFA is critical to be able to deliver bunches with low emittance and energy spread after multiple stages of acceleration.
Optimizing the beam loading of the wake is crucial for controlling the energy spread such that all particles throughout the bunch experience the same accelerating gradient~\cite{Katsouleas_PA87, Chen_PRL86, Lotov_PoP05, Tzoufras_PRL08}. Emittance preservation largely depends on proper matching~\cite{Xu_RPL16} and alignment~\cite{Thevenet_PRAB19} of the trailing bunch within the plasma bubble, and preserving this matching within the entrance and exit ramps of the plasma~\cite{Ariniello_PRAB19, Ariniello_PRR22}. 
Additional factors such as ion motion~\cite{An_PRL17} and beam scattering~\cite{Zhao_PoP20} can also have deleterious effects on emittance.

Particle-in-Cell (PIC) simulations play a vital role in considering all of these parameters and determining PWFA schemes that optimize both energy transfer efficiency and preservation of beam quality.
Previous publications have demonstrated a strategy for achieving $>\SI{10}{\giga\electronvolt}$ energy gain with high efficiency, preservation of energy spread and emittance of a \SI{0.5}{\nano\coulomb}, \SI{10}{\micro\meter} emittance trailing bunch with the ultimate beam parameters that will be available at FACET-II~\cite{Joshi_2018}.

\begin{figure}[htp]
    \centering
    \includegraphics[width=0.5\textwidth]{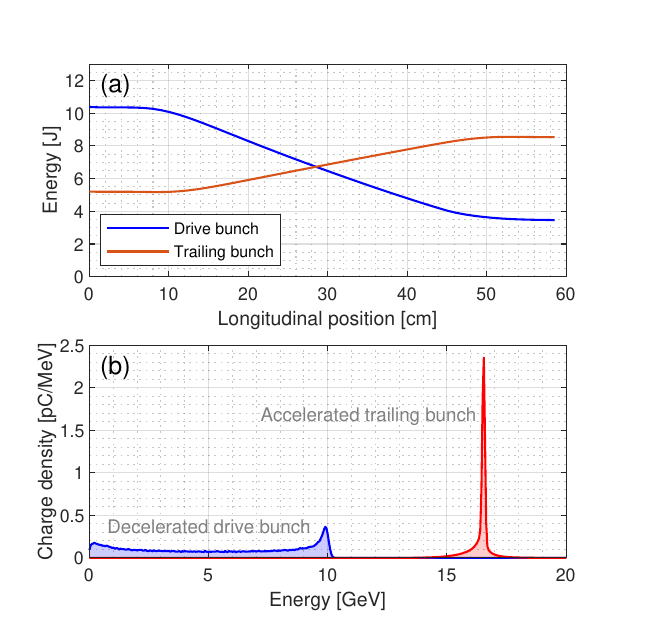}
    \caption{PIC simulation of PWFA performance with initial FACET-II beam parameters, including incoming beam energy of \SI{10}{\giga\electronvolt} for both drive and trailing bunches. (a) the evolution of the total energy content of the drive and trailing bunches as they traverse a lithium plasma with \SI{40}{\centi\m} long flat-top profile at a density of \SI{8e16}{\centi\meter^{-3}}, resulting in an overall drive to trailing bunch efficiency of 32\%.
    (b) shows the final energy spectra of the drive and trailing bunches, showing an acceleration of the trailing bunch by \SI{6.6}{\giga\electronvolt} with final energy spread of 0.9\%.}
    \label{fig:pic}
\end{figure}

In the initial phases of beam development, FACET-II will be operating with the relaxed beam parameters which will be discussed in the following sections.
PIC simulations using QPAD~\cite{QPAD} have been performed to provide insights into the expected performance of PWFA with these conditions, see FIG. \ref{fig:pic}.
In the simulation, a moving window with dimensions of $z=\SI{225}{\micro\meter}$ (beam direction) and $r=\SI{168}{\micro\meter}$ (transverse direction) was used. The simulation box was divided into 1600 and 400 cells along the $z$ and $r$ directions, respectively. In the azimuthal direction, 16 cells were used for both bunches and the plasma.  16 particles were initialized in each cell for both drive and trailing bunches, and 64 particles per cell was used for the plasma.
As the drive and trailing bunches were modeled as Gaussian bunches in this simulation, only the $m=0$ (lowest order) mode was included.
The plasma source is modeled as a neutral lithium gas with \SI{40}{\centi\m} long flat-top plasma density of \SI{8e16}{cm^{-3}} and realistic entrance and exit density ramps, and beam-ionization calculated using the ADK model~\cite{ADK}.

In the two-bunch mode of beam delivery, where the linac delivers both the drive and trailing bunch, this assumes a \SI{1.5}{\nano\coulomb} drive bunch with \SI{25}{\micro\meter} emittance and \SI{27}{kA} peak current, and a \SI{0.5}{\nano\coulomb} trailing bunch with \SI{30}{\micro\meter} emittance and \SI{7}{kA} peak current.
With these parameters, simulation results show that the drive bunch will approach energy depletion with 66\% drive-to-wake transfer efficiency.
Furthermore, a trailing bunch separated by \SI{106}{\micro\meter} can be accelerated by \SI{6.6}{\giga\electronvolt} with wake-to-trailing bunch efficiency of 48\%.

The drive bunch transfers \SI{6.9}{J} out of the initial \SI{10.4}{J} of energy into the wake, while the trailing bunch picks up \SI{3.3}{J}, for an overall drive to trailing bunch efficiency of 32\%. The energy spectrum of the two bunches after traversing through the plasma shows the large energy spread of the energy-depleted drive bunch extending to near zero energy, and the trailing bunch accelerated with energy spread remaining below 1\%.
The emittance of the trailing bunch is preserved at the incoming \SI{30}{\micro\meter} throughout the simulation.

\section{Experimental Setup}

The FACET-II facility will provide electron bunches for beam-driven PWFA studies with energy of \SI{10}{\giga\electronvolt} in either single or two bunch configurations~\cite{Yakimenko_PRAB19}.
Significant modifications have been made to the \SI{2}{km} SLAC linac previously used for FACET for PWFA studies between 2012 and 2016.
These include the removal of the initial \SI{1}{km} segment of the original SLAC linac, which housed the injector, damping rings, and accelerating structures, to accommodate the installation of the new LCLS-II (Linac Coherent Light Source) superconducting RF accelerator.
A new photocathode injector has been installed to generate beams with smaller and symmetric emittances, capable of producing either single or two bunches directly from the cathode.
The linac now also contains three stages of bunch compression, enabling compression to peak currents exceeding \SI{100}{kA}.
Although positron capabilities are not currently available due to the removal of the original SLAC damping rings, future plans involve the reinstatement of this capability.
This will be achieved by installing a new compact positron damping ring and insertion beamline, allowing for the simultaneous delivery of electron and positron bunches for PWFA studies~\cite{FACET_TDR}.

The flexibility of the FACET-II accelerator allows it to be operated in several different configurations to meet the needs of the various experimental programs. In the two bunch delivery mode, a double laser pulse on the cathode of the RF photocathode injector will produce the two bunches -- \textit{drive} and \textit{trailing}, that are co-accelerated through the linac.After the final compression, the drive and trailing bunches will be separated longitudinally by $\sim$\SI{150}{\micro\meter}.
At the time of writing, the accelerator has been commissioned in the single bunch configuration, delivering bunch charges of up to \SI{2}{nC} with \SI{20}{\micro\meter} normalized emittance.
The addition of a laser heater in the FACET-II injector adds further longitudinal phase space control and suppression of the microbunching instability~\cite{Huang_PRAB10}.
The design beam parameters of both drive and trailing bunches in this mode of operation are listed in Table \ref{tab:FACET}, along with the currently achieved beam parameters.

\begin{table}
\caption{\label{tab:FACET}FACET-II beam parameters for two-bunch PWFA. The two bunch parameters are listed as drive/trailing.}
\begin{ruledtabular}
\begin{tabular}{lcc}
Electron beam parameter & Current\footnote{Parameters achieved at time of preparation of this manuscript} & Design \\
\colrule
Bunch configuration & Single & Two-bunch \\
Delivered beam energy, (GeV) & 10 & 10.1 / 9.9\\
Norm. emittance, (mm-mrad) & $\sim20$ & $>$ 50 / 5\ \\
Charge per bunch, (nC) & 2 & 1.5 / 0.5 \\
Peak current, (kA) & -- &  30 / 15 \\
RMS energy spread, (\%) & $\sim1$ & 0.8 / 0.3 \\
Repetition rate, (Hz) & 1 -- 30 & 1 -- 30 \\
IP $\beta^{*}$, (cm) & 50 & 5 -- 50 \\ 
\end{tabular}
\end{ruledtabular}
\end{table}

\begin{figure*}[htb]
    \centering
    \includegraphics[width=\textwidth]{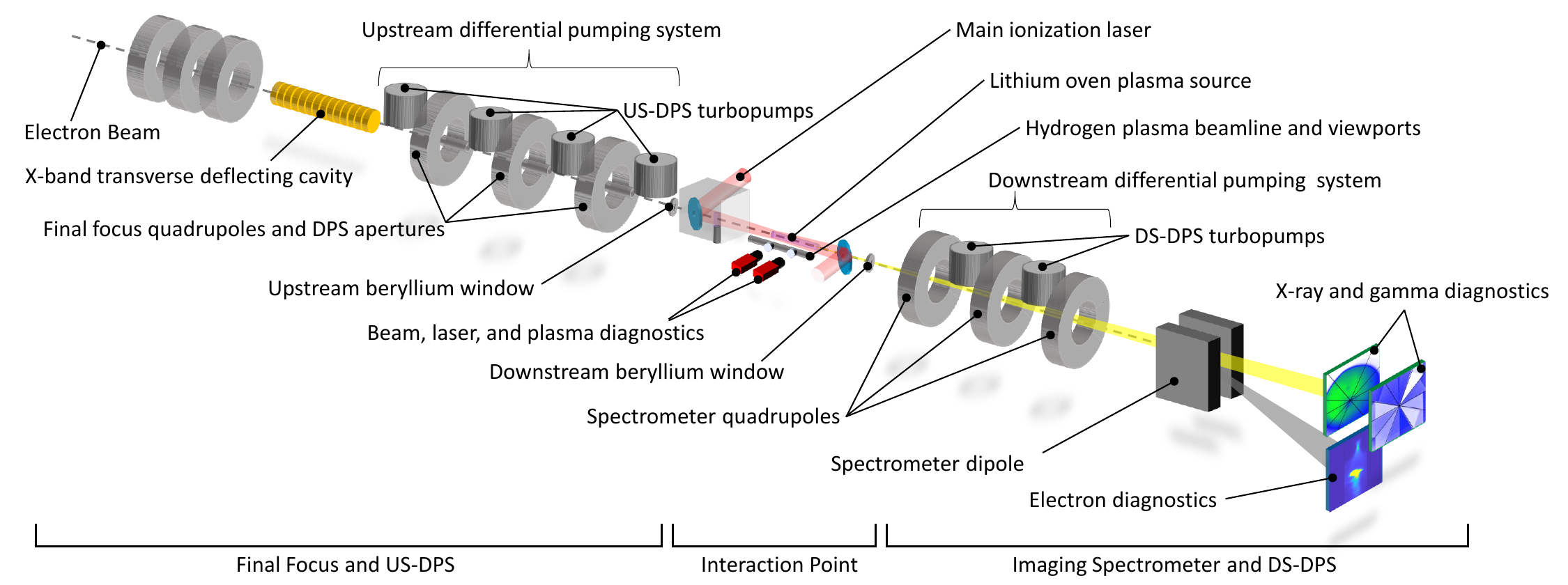}
    \caption{The FACET-II experimental area beamline showing the key hardware in the plasma research program.  The electron beam travels from left to right in the schematic. The final focus system is on the left, containing the X-band transverse deflecting cavity and upstream differential pumping system (US-DPS).
    This is followed by the Interaction Point (IP) area containing the plasma sources, laser integration optics, and diagnostics. The lithium oven and hydrogen plasma beamline may be remotely actuated to switch the beam path between plasma sources.
    The spectrometer beamline transmits the beam to the beam dump and contains the downstream differential pumping system (DS-DPS) and electron and betatron diagnostics.}
    \label{fig:DPS}
\end{figure*}

After acceleration and compression, the beam is delivered to the experimental area, which has been designed to simultaneously accommodate a wide range of experiments including advanced acceleration techniques such as PWFA, applications of machine learning for accelerator diagnostics and control, novel techniques for the generation of intense coherent radiation, and probing strong-field quantum electrodynamics~\cite{Storey_IPAC23}.  A schematic of the experimental area is shown in FIG. \ref{fig:DPS}, highlighting the key components relating to the PWFA research program.

The final focusing system consists of two quadrupole triplets capable of focusing the few \SI{}{\micro\m} emittance beams to a 3--4\,\SI{}{\micro\m} spot size at the beam waist, allowing for matching into the plasma source.
The Interaction Point (IP) area contains the plasma sources, laser integration optics, d electron and laser diagnostics for the wide-ranging experimental program.
This is followed by an imaging electron spectrometer which consists of a magnetic quadrupole triplet for capturing and refocusing the beam exiting the IP, and a dipole magnet to provide vertical dispersion for energy resolved measurements. These diagnostics will be described in detail in section \ref{sec:diag}.

\subsection{\label{sec:oven}Plasma sources}

Two types of plasma sources are used at the IP for PWFA studies, using either lithium vapor or hydrogen gas to form a plasma. Lithium plasma sources have been employed in the previous PWFA experiments at SLAC at the Final Focus Test Beam (FFTB) and FACET facilities. Using lithium for the plasma source makes use of the fact that the outermost electron of lithium is relatively easy to ionize either by beam-ionization or by preionization with a laser. Hydrogen gas may be used in either a static fill or gas jet, and has a higher ionization threshold which allows for plasma ramp shaping via laser ionization~\cite{Ariniello_PRAB19}.

The lithium plasma is generated in a heat pipe oven where a uniform column of neutral lithium vapor is produced by heating a section of beam pipe containing solid lithium to temperatures of up to \SI{1000}{\celsius}. The uniform density region of this column is contained to a length of approximately \SI{40}{\centi\meter} by a helium buffer gas exerting several Torr of pressure. This results in (10\%-90\%) lithium density ramps at the start and end of the heated section over a length of approximately \SI{10}{\centi\meter}~\cite{Navid_AAC12}.
Although the vapor pressure of lithium is very sensitive to the oven temperature, we have achieved near uniform flat top lithium vapor column densities of up to \SI{8e16}{cm^{-3}} with helium buffer pressures of up to \SI{10}{Torr}.
The low ionization energy of lithium compared to helium allows for the formation of a pure lithium plasma with number density matching the gas density of lithium when field-induced beam ionization is used.
For small incoming emittance, partial beam-ionization of the helium buffer gas can occur if the beam density reaches the helium ionization threshold, disrupting matching into the plasma ramps and injecting dark current into the plasma~\cite{Navid_PRL14}. 

A parallel beamline is positioned adjacent to the lithium plasma oven to allow for beam tuning without passing the beam through the lithium oven when it is at operational temperature. The oven and this bypass beamline are installed on an actuated table to allow for remotely switching between the two. The bypass beamline also allows for the beam to be passed through a static fill of gas, such as hydrogen, that can be ionized by either the beam or a laser.

The maximum repetition rate that a plasma accelerator may be operated at is determined by the time interval for the plasma to \textit{reset} between shots, through recombination and diffusion of the ionized atoms, and the return to thermal equilibrium.
The energy deposited into the plasma by the drive beam will go into increasing the overall thermal-kinetic energy of the system.
For the lithium vapor plasma oven, this increased energy will act to lengthen the gas column as the more energetic atoms push outwards on the buffer gas at constant pressure.
This will move the location of the lithium entrance and exit plasma ramps outwards, affecting both the matching conditions and length of the plasma.
This places a limit on the repetition rate on the order of \SI{1}{Hz} when operating continuously, or 10 Hz for shorter bursts.
In a steady state system, this effect may be counteracted by a feed-forward system that adjusts the oven heater power to match the energy deposited into the plasma by the beam.
However, this can be difficult to achieve in a research setting where the energy deposited into the plasma is difficult to predict due to varying input parameters.
Alternatively, a meter-long  hydrogen gas flowing out of multiple overlapping supersonic nozzles may allow much higher repetition rate operation, at least in a burst mode, by the complete replacement of the gas in the plasma interaction region between shots.

\subsection{\label{sec:DPS}Differential Pumping System}

\begin{table*}[btp]
\caption{\label{tab:DPSresults}Beamline pressures in Torr under several operating modes of the differential pumping system. The numbering of the stages increases with distance from the IP. The beamline pressure is reduced to the baseline vacuum pressure at the XTCAV location and to well below the spectrometer vacuum requirement in all modes of operation.}
\begin{ruledtabular}
\begin{tabular}{lc ccccc ccc}
 & & \multicolumn{5}{l}{Upstream (US) stages:} & \multicolumn{3}{l}{Downstream (DS) stages:} \\
 Mode &  IP \ \  & US1 & US2 & US3 & US4 & XTCAV \ \ & DS1 & DS2 & Spectrometer \\
\colrule
Baseline & 0 & \SI{6e-10}{} & \SI{1e-8}{} & \SI{2e-9}{} & \SI{1e-10}{} & \SI{3e-9}{} & \SI{5e-9}{} & \SI{1e-9}{} & \SI{2e-8}{} \\
5 Torr He & 5.0 & \SI{3e-3}{} & \SI{5e-6}{} & \SI{2e-8}{} & \SI{1e-9}{} & \SI{3e-9}{} & \SI{2e-4}{} & \SI{1e-6}{} & \SI{5e-7}{} \\
10 Torr He & 10.0 & \SI{7e-3}{} & \SI{1e-5}{} & - -\footnote{Gauge inoperable during the test}  & \SI{1e-9}{} & \SI{3e-9}{} & \SI{5e-4}{} & \SI{7e-6}{} & \SI{8e-7}{} \\
5 Torr H$_2$ & 5.0 & \SI{1e-2}{} & \SI{2e-5}{} & \SI{2e-8}{} & \SI{3e-9}{} & - -\footnote{Isolated by beamline valve during the test} & \SI{1e-3}{} & \SI{5e-6}{} & \SI{1e-7}{} \\
5 Torr Ar & 5.0 & \SI{1e-3}{} & \SI{5e-6}{} & \SI{5e-9}{} & \SI{2e-10}{} & \SI{3e-9}{} & \SI{6e-5}{} & \SI{5e-7}{} & \SI{4e-8}{} \\
\end{tabular}
\end{ruledtabular}
\end{table*}

In the prior implementation at FACET, the experimental area was separated from the rest of the linac by a 50--\SI{75}{\micro\m} thick beryllium window to contain the gas associated with the plasma sources.
However, at the increased beam intensities of FACET-II, any solid material in the beam path near the IP will be destroyed by beam heating~\cite{Stupakov}.
At the ultimate FACET-II beam parameters, the solid windows will be destroyed within a single shot by this effect. A differential pumping system (DPS) has therefore been implemented to reduce the vacuum pressure along the beamline without the need for solid windows in the beam path.
Additionally, the DPS has the added benefit of being able to deliver the lowest emittance and highest charge (brightest) beams to the IP that the photocathode gun enabled linac can deliver.

The DPS is comprised of upstream and downstream systems (US-DPS and DS-DPS), located on either side of the IP area as depicted in FIG. \ref{fig:DPS}.
Each side contains a series of pumping stages interleaved with the final focus and spectrometer quadrupoles along the beamline, and separated by conductance limiting beam-pipes, which decrease the pressure by several orders of magnitude at each stage.

On the upstream side, four stages of differential pumping decrease the beamline pressure from up to \SI{10}{Torr} at the IP down to approximately \SI{1}{nTorr} at the location of the X-band RF transverse deflecting cavity (XTCAV)--located approximately \SI{6}{m} upstream of the lithium oven.
This is required to avoid RF breakdown and structure damage during operation of the XTCAV.
On the downstream side of the IP, the beamline pressure must be reduced to $<\SI{1}{mTorr}$ to limit gas scattering in the spectrometer beamline which can contribute to emittance growth and degradation of beam measurement quality.
This is accomplished by two stages of differential pumping.
Each stage of pumping is provided by a magnetically levitated turbomolecular pump mounted directly to the beamline. The turbopumps were chosen for their high pumping speeds for both helium and hydrogen and for their use of magnetic levitation bearings to limit vibration transfer to the beamline.

Initial conductance restricting apertures are inserted into the beamline on either side of the IP to achieve the first pressure drop from Torr-level pressures down to several mTorr in the first stage of differential pumping. Currently this first aperture restriction is simply the previously installed beryllium windows which have had $\sim\SI{200}{\micro\meter}$ diameter holes drilled through in-situ by the high intensity electron beam. This has ensured that the holes are self-aligned to the nominal electron beam path without further alignment intervention.
While this small orifice-type aperture is sufficient for initial commissioning, these will eventually be replaced with straw-type apertures with \SI{5}{mm} diameter and \SI{100}{mm} length to limit background and emittance growth from beam halo scattering on the edges of the holes.
The apertures on either side of the IP are separated by a distance of approximately \SI{4}{\meter}.
The beam pipes within the quadrupole magnets provide sufficient conductance limitation between the later stages in the US-DPS and DS-DPS to reach the required reduction in beamline pressure.

The DPS has been commissioned in various modes of operation including \textit{static} fill of either helium or argon up to IP pressures of \SI{10}{Torr}, or hydrogen gas up to 5 Torr pressure at the IP. The system has also been used with high pressure hydrogen and helium gas-jet plasma sources operating at the IP with up to \SI{10}{Hz} repetition rate. The vacuum pressures at each stage of the DPS in static fill operations are listed in Table \ref{tab:DPSresults}, showing the pressure dropping to the mTorr pressure level on the first stages on either side of the IP, and reaching down to $\sim$nTorr pressures at the location of the upstream XTCAV, and $<\SI{1e-6}{Torr}$ in the spectrometer beamline.

A full demonstration of the first lithium oven operation with the differential pumping system has been performed, with the oven operating with 5 Torr helium buffer gas pressure for $>24$ hours. The buffer gas pressure was maintained to within $\pm2\%$ for the duration of the test, with a further upgrade reducing the pressure stability to a diurnal variation of $<0.5\%$.

\section{\label{sec:diag}PWFA Diagnostics}

The experimental area is equipped with a variety of diagnostics that are used to quantify the performance of the plasma wakefield acceleration.
The primary electron diagnostic is an imaging spectrometer beamline that captures, refocuses, and vertically disperses the electron beam after the plasma.
This allows for energy-resolved measurements on profile monitors located approximately \SI{20}{\meter} downstream of the IP, just before the beam dump.

A set of multipurpose spectrometer diagnostics have been designed to meet the needs of various user programs that use FACET-II. Electron diagnostics include a high resolution in-vacuum profile monitor located immediately prior to the vacuum exit window, a large field of view camera that images a  gadolinium-oxysulfide (GOS) scintillator screen just after the exit window, and a Cherenkov light spectrometer~\cite{Adli_NIMA15} that images the dispersed electron beam from full energy down to $\sim\SI{1}{\giga\electronvolt}$.
A series of photon imaging screens on the zero-dispersion axis image the X-rays and gamma rays generated by the electron beam's betatron motion in the plasma, providing intensity, transverse, and spectral information.

FACET-II is capable of producing beams with exceptionally high current and beam density, resulting in extremely challenging conditions for intercepting diagnostics near the IP where the beam density and strong fields can destroy any solid material within a single shot~\cite{Green_IBIC17}.
While standard optical transition radiation (OTR) screens and wire scanners are used in the IP area, their use is limited by beam conditions. Non-intercepting diagnostics, which will be described in the following sections, are employed whenever possible.

An X-band transverse deflecting cavity  is located before the plasma source to measure the incoming longitudinal profile of the beam, using an imaging screen in the spectrometer beamline.
A probe laser system allows for non-invasive electro-optical sampling (EOS) for both longitudinal and transverse measurements of the incoming beam.
The OTR and laser diagnostic cameras that view the beamline through viewports along the oven bypass line may also be repurposed for the direct visualization of the plasma emission light from hydrogen plasma at discrete locations. 
In addition, a series of thermocouples monitor the temperature distribution of the lithium oven to ensure the reproducibility of the lithium vapor column.

The measurements enabled by these diagnostics are summarized in the following sections.

\subsection{\label{sec:emit}Emittance Measurements}

Precision measurements of beam emittance both before and after the beam undergoes acceleration are essential for demonstrating the preservation of beam quality in PWFA.
Standard methods of measuring emittance involve measuring the beam size either in multiple locations separated by a set of known optical elements in the multi-screen method, or at a single location in a multi-shot quadrupole scan measurement.  The former method requires a significant length of beamline to implement multiple sets of transverse diagnostics and magnetic elements, while the latter requires multiple shots to be acquired over a time period of at least 10s of seconds while quadrupole strengths are changed.
Both methods are invasive and cannot be carried out simultaneously with beam energy measurements using the spectrometer.
A single-shot measurement of the horizontal emittance can be made in a single plane by analysing the transverse profile of the beam in a magnetic spectrometer beamline~\cite{Wiengartner_PRSTAB12, Barber_APL20}.  This type of measurement was previously used in FACET to provide an upper limit to emittance measurements of beam-driven PWFA accelerated beams~\cite{Navid_PPCF16}.

The transverse beam size at the \textit{image plane} in the spectrometer, $\sigma_x$, can be related to the normalized emittance, $\epsilon_n$, and Twiss parameters $\beta_0$ and $\alpha_0$ at the \textit{object plane}, i.e. at the exit of the plasma.  Using the known transport matrix, $M_{ij}$, through the spectrometer beamline from the object plane to image plane, the horizontal spot size for particles of a particular energy (and Lorentz factor $\gamma$) is given by the formula

\begin{equation}
    \label{eq:emit}
    \sigma_{x}(E)^2 = \frac{\epsilon_n}{\gamma}\left[  M_{11}^2\beta_0 - 2M_{11}M_{12}\alpha_0 + M_{12}^2\left( \frac{1+\alpha_0^2}{\beta_0} \right) \right]
\end{equation}

\noindent where $M_{11}$ is the relation between positions at the object plane and the image plane, and $M_{12}$ is the relation between the angle at the object plane and position in the image plane. 

In the imaging condition, the spectrometer quadrupoles are set such that the transport matrix provides point-to-point imaging between the object and image planes, with the matrix element $M_{12} = M_{34} = 0$ for particles at the energy set-point of the spectrometer.
The value of $M_{12}$ becomes non-zero for electron energies away from the energy set-point due to the chromaticity of the spectrometer.
Therefore, if the beam is at waist at the object plane, then the beam size will be re-imaged to the smallest spot size for particles at the setpoint energy where the imaging condition is perfectly met, and increase in size for electrons with larger and smaller energy.
With the spectrometer dipole deflecting the beam vertically, the dispersion at the image plane results in an hourglass or butterfly-shaped transverse beam profile, allowing for the beam width to be extracted as a function of energy. Operating with $M_{34}=0$ ensures the highest possible energy resolution.

By fitting the measured horizontal beam size as a function of energy to equation (\ref{eq:emit}) using the known dependency of the transport matrix elements with energy, the horizontal projected emittance and Twiss parameters at the object plane can be extracted.
This analysis holds in the condition that these parameters do not vary with energy, so is constrained by chromatic correlations (i.e. $x$-$E$) and phase mismatch within the plasma, which cause variations in emittance and Twiss parameters that are dependent on energy.
This technique does not however require perfect knowledge of the plasma exit position, as this may be extracted via the fitted Twiss parameters, providing useful information into the true plasma exit location.

\begin{figure}[thb]
    \centering
    \includegraphics[width=0.5\textwidth]{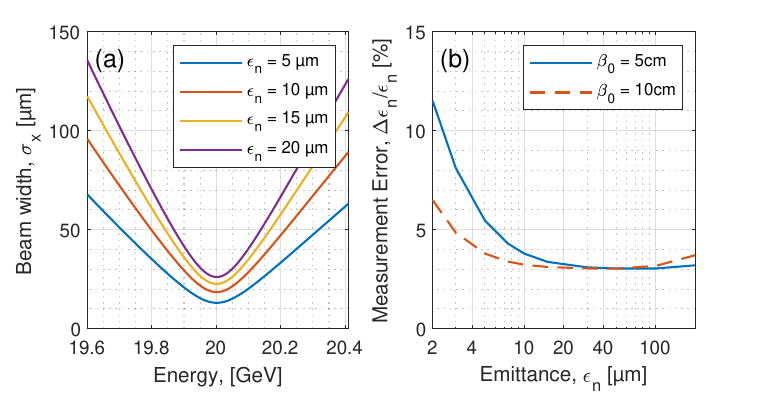}
    \caption{(a) The electron beam size that would be measured at the spectrometer image plane for several different values of emittance with $\beta_0=\SI{5}{\centi\m}$ and the spectrometer set to image \SI{20}{\giga\electronvolt} electrons.
    (b) an estimate of the emittance measurement error for an energy-doubled trailing bunch with 0.5\% energy spread, accounting for the beam transport properties and an optical imaging resolution.}
    \label{fig:BFtheory}
\end{figure}

For the FACET-II beam parameters, the high energy of the accelerated trailing bunch and low transverse emittance result in transverse beam profiles with \SI{}{\micro\m}-size features.  FIG. \ref{fig:BFtheory} shows the horizontal beam size at the spectrometer image plane as a function of energy for several emittances for a fixed vacuum $\beta_0=\SI{5}{\centi\meter}$ at the exit of the plasma oven, showing a minimum spot size on the order of \SI{10}{\micro\m}. The magnification of the spectrometer when imaging the plasma exit plane at \SI{20}{\giga\electronvolt} is $M_{11}=5.2$.
By employing a high resolution in-vacuum OTR beam profile monitor with optical imaging resolution of \SI{4.5}{\micro\m}, the emittance can be extracted from single-shot images of matched beams at the plasma exit with several-\% level measurement uncertainty for beam emittances greater than $\simeq$\SI{10}{\micro\m}. This uncertainty estimate accounts for imaging resolution, uncertainty in the transport matrix elements, and signal noise in the imaging system.

Alternatively, if the conditions are stable enough to allow for a multi-shot measurement, a dispersive quadrupole scan can instead be made to extract the emittance for multiple energy slices across the beam. 
In this measurement, the quadrupole strengths are scanned over a range to vary the transport matrix while acquiring images of the energy dispersed transverse beam profile at each setting.
Fitting the horizontal spot size at a given energy over the range of $M_{11}$ and $M_{12}$ using equation (\ref{eq:emit}) provides the emittance and Twiss parameters at the object plane as a function of energy.
This type of measurement allows for an assessment of the variation in beam parameters across the bunch, and is crucial for measuring the incoming trailing beam parameters when the energy spread is too small to allow for enough statistics for the single-shot emittance measurement.
This allows for accurate emittance measurements for mismatched beams, where the assumption that emittance and Twiss parameters are constant across the bunch does not hold.
The main limitation of this measurement is that it requires stable beam conditions over the time scale it takes to vary the quadrupole strengths and acquire imaging data at each point. This is typically on the order of several minutes for a single scan.

\begin{figure}[th]
    \centering
    \includegraphics[width=0.5\textwidth]{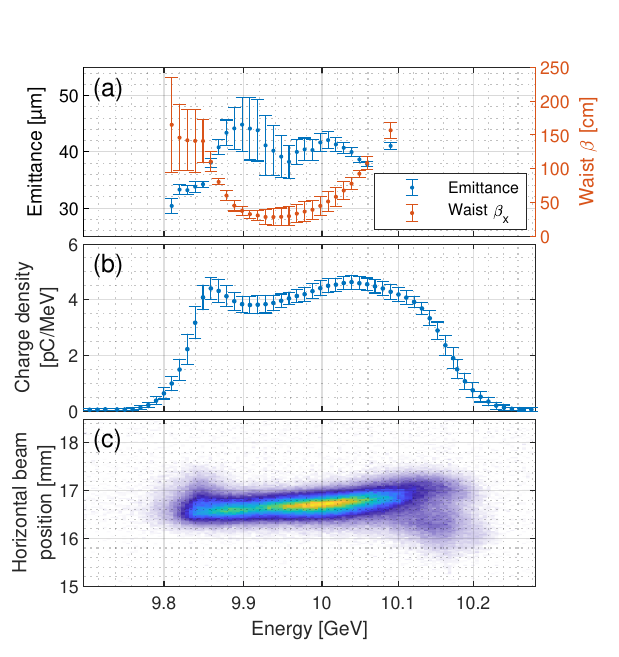}
    \caption{The measured dispersive quadrupole scan of the beam in the single-bunch configuration. The plot in (a) shows the emittance extracted as a function of energy, with a value of approximately \SI{40}{\micro\m} across the core of the bunch, and the waist $\beta$ with a minimum value of \SI{30}{\centi\m} in this range. The beam profile becomes non-Gaussian at energies above \SI{10.1}{\giga\electronvolt}, preventing us from accurately reconstructing the emittance and $\beta$ values in this range.  (b) shows the current profile of the bunch, and (c) shows a single-shot image of the beam in the middle of the quadrupole scan when $M_{12}$ is nominally set to 0.}
    \label{fig:QS}
\end{figure}

Commissioning of the emittance diagnostics has been performed using the single-bunch configuration with emittance on the order of 20-\SI{40}{\micro\m} and 1\% energy spread.  The final focus and spectrometer optics were configured to deliver the beam with $\beta_0=\SI{50}{cm}$ at the location of a wire scanner, just prior to the nominal plasma entrance location, and re-imaged to the high resolution spectrometer beam profile monitor. FIG. \ref{fig:QS}(a) demonstrates a dispersive quadrupole scan measurement, indicating an emittance of approximately \SI{40}{\micro\meter} measured across the core of the bunch and a waist $\beta$ of \SI{30}{\centi\m}.  This tool will be instrumental to diagnosing the matching conditions of the drive and trailing bunches incoming to the plasma.

In the present beam configuration, the Twiss parameters vary too significantly across the bunch to allow for a single-shot emittance measurement of the incoming single-bunch beam. In the two-bunch configuration, the smaller energy spread of the trailing bunch leads to reduced parameter variation across the bunch, enabling the single-shot measurements to be applied more effectively.
Nevertheless, this measurement has already been applied to data acquired during the initial PWFA commissioning, and its findings will be discussed in Section \ref{sec:beamtests}.

\subsection{\label{sec:energy}Electron Energy Spectra}

The beam exiting the plasma ranges in energy from less than \SI{1}{GeV} in the energy-depleted drive bunch to more than \SI{20}{GeV} in an energy-doubled trailing bunch. Measurement of the full energy range of the drive bunch is important to understand the driver energy depletion and drive-to-wake transfer efficiency. On the other hand, high resolution measurements of the trailing bunch are required to demonstrate preservation of the energy spread at the level of $<1\%$. The electron spectrometer beamline therefore uses different screens to meet these differing needs.

The trailing bunch will be measured by the high resolution transverse beam profile monitor used for the emittance measurements. The energy resolution can be determined by adding in quadrature both the imaging resolution $\sigma_{im}$, and the transverse beam size as determined by the vertical emittance, $\epsilon_n$, and $\beta$-function: $\sigma_{R,res}/E = \sqrt{\sigma_{im}^2+\beta_y\epsilon_n\gamma^{-1}}/\eta$. The vertical dispersion, $\eta$, at the location of the screen is nominally \SI{60}{mm}, and the imaging resolution was measured to be \SI{4.5}{\micro\m}. $\gamma$ here again refers to the Lorentz factor at the measurement energy $E$.

In the nominal PWFA configuration, the transverse beam size at the imaging plane $\sqrt{\beta_y\epsilon_n\gamma^{-1}}$ will be ${<\SI{10}{\micro\m}}$ for an emittance preserved beam, leading to an overall energy resolution of $<0.02\%$ ($\sim\SI{2}{\mega\electronvolt})$.
For the initial run parameters with normalized emittance on the order of \SI{40}{\micro\meter} and with an IP beta function of \SI{50}{cm}, the energy resolution at \SI{10}{GeV} is dominated by the transverse beam size of $\sim$\SI{50}{\micro\m}, leading to a resolution of 0.1\% (\SI{10}{\mega\electronvolt}). The energy profile of the current single bunch beam without plasma interaction was shown in FIG. \ref{fig:QS}(b and c), with a measured FWHM energy spread of 3\%.

To measure large electron energy ranges, either of two large field of view profile monitors are used. The first employs the GOS-based scintillator, DRZ\texttrademark-FINE ~\footnote{Manufactured by Mitsubishi Chemical Group}, that stretches from above the zero-dispersion axis down to a dispersion level of \SI{120}{mm}.
At the nominal setting of the spectrometer dipole, the field of view extends down to an energy of approximately \SI{5}{\giga\electronvolt}. The spectrometer dipole strength may be lowered down to 25\% of the nominal value, decreasing the lower extent visible on this diagnostic down to $\sim\SI{1}{GeV}$. 
The relative energy resolution of this diagnostic is dominated by the pixel size of the imaging system, resulting in an energy resolution of 0.15\%.
A second screen located several meters upstream can be used to extend the low energy portion of the spectrum to $\sim\SI{0.25}{\giga\electronvolt}$.
Charge below \SI{250}{\mega\electronvolt} will be undetectable by direct observation using the magnetic  spectrometer as these low energy electrons are deflected into the wall of the spectrometer dipole chamber prior to the opportunity for their measurement.
The dipole strength may also be increased to allow the high energy portion of the electron spectrum to be imaged with higher energy resolution as an energy gain diagnostic.

The main limitations of the scintillation screen spectrometer diagnostic are the saturation of the scintillating centers and damage to the screen material leading to permanent loss of light output at locations of high beam intensity.
To overcome these challenges, the second large field of view diagnostic that is employed is a Cherenkov light-based electron spectrometer that is described in detail in ~\cite{Adli_NIMA15}. This transverse beam profile monitor images the Cherenkov light emitted by beam electrons as they pass through a small air gap before the beam dump.
The Cherenkov light is reflected from the beam path by a beam intersecting polished silicon wafer.
Since Cherenkov light is emitted with high linearity with charge density, and the silicon reflecting surface has a relatively high damage threshold, this diagnostic provides a higher dynamic range and robustness than the scintillator screen.
The spatial resolution at \SI{10}{GeV} is \SI{250}{\micro\meter}, limited mainly by the multiple scattering of the beam as it passes through the \SI{5}{mm} aluminum vacuum exit window. This translates to an energy resolution of 0.4\% at 10 GeV.

Additional non-invasive measurements of the energy spectrum of the incoming beam are performed using a synchrotron light diagnostic (SYAG) prior to the IP. This device is located within the final bunch compressor at a location with large horizontal dispersion and is comprised of a short, three-magnet vertical chicane to generate the emission of synchrotron photons from the beam electrons which are intercepted by a Cerium doped Yittrium Aluminum Garnet (YAG) scintillator screen for detection.
Due to the horizontal dispersion of the beam at this location, the horizontal profile of the X-rays represents the energy distribution of the electron beam.

During the initial phases of beam development, the presence of coherent OTR (COTR) detected on OTR screens near the IP and spectrometer diagnostics indicates that the electron bunches can contain high current structure on top of the bunch profile visible on the present diagnostics.
The source of this high current structure is possibly due to unmitigated microbunching occurring early in the linac.
While the microbunching itself will ultimately be suppressed by the use of the laser heater, the SYAG diagnostic can provide some non-intercepting information about the presence of this longitudinal structure due to the energy chirp on beam.

\begin{figure}[th]
    \centering
    \includegraphics[width=0.5\textwidth]{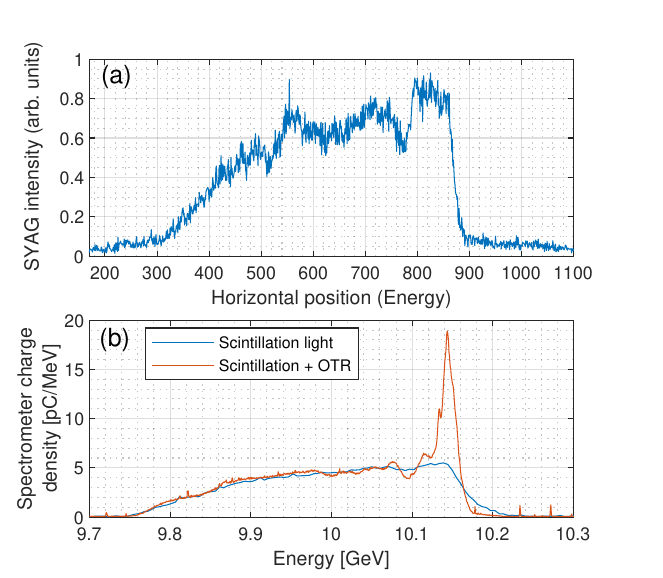}
    \caption{The SYAG screen intensity measured on SYAG in (a), and the energy spectrum measured on the in-vacuum electron spectrometer in (b) using two cameras simultaneously viewing a transparent YAG crystal oriented at \SI{45}{\degree} to the beam direction. The camera viewing the back side of the YAG sees only scintillation light, and provides a consistent measurement of total charge.  Meanwhile, the camera viewing at the angle of OTR emission images a strong coherent OTR emission at the head of the bunch for the same shot.}
    \label{fig:SYAG}
\end{figure}

FIG. \ref{fig:SYAG} shows the profile measured with the SYAG diagnostic for a single shot that resulted in significant COTR emission, and the corresponding energy spectrum measured with the in-vacuum electron spectrometer using two cameras viewing the same YAG screen.
One camera views the back surface of the YAG crystal oriented at \SI{45}{\degree} to the beam. In this orientation, the camera only collects the scintillation light, and not the forward emitted COTR.
While the resolution of this camera is limited to 10s of \SI{}{\micro\meter} due to the thickness of the YAG crystal, the total intensity measured is consistent with the bunch charge, and the profile matches the SYAG profile, within the resolution limits.
A second higher resolution camera views the front surface of the same YAG screen. In this orientation, the camera collects both scintillation light and the backwards OTR emission. In this shot, this camera sees a large non-linear intensity spike at the head of the bunch due to the emission of COTR.
While the SYAG diagnostic does not have the resolution to fully resolve the fine structure that leads to coherence in OTR in the optical range, it can be used to monitor for shot-to-shot longitudinal variations that often accompany the microbunching instability.

\subsection{\label{sec:LPS}Incoming Longitudinal Phase Space}

\begin{figure}[th]
    \centering
    \includegraphics[width=0.48\textwidth]{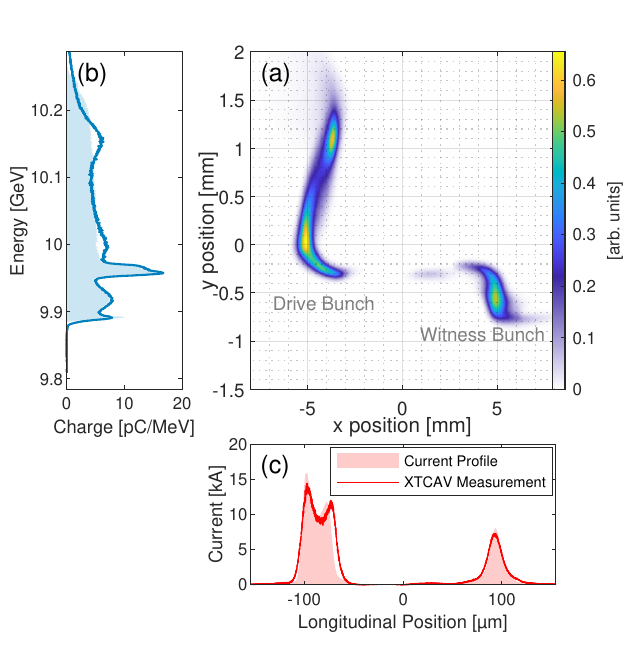}
    \caption{Simulated measurement of the longitudinal phase space measured with the XTCAV and the high resolution beam profile monitor in the magnetic spectrometer.
    (a) The transverse profile imaged on the spectrometer high resolution beam profile monitor. The horizontal position axis provides the longitudinal profile due to the horizontal streak applied by the XTCAV and the vertical position axis provides the energy due to the magnetic spectrometer dipole.
    In (b) and (c), the beam energy and current profiles are shown as the shaded areas and the measured profiles from the calibrated XTCAV image are represented by the solid lines.
    With the beamline optics tuned to the trailing bunch energy, the drive bunch reconstruction suffers slightly due to the energy difference.}
    \label{fig:XTCAV}
\end{figure}

The X-band transverse deflecting cavity~\cite{Dolgashev_PRAB14} (XTCAV) allows for longitudinal profile measurements with down to \SI{1}{\micro\meter} ($\sim\SI{3}{fs}$) resolution. In the reconfiguration of the experimental area for FACET-II, this structure has been relocated to within the final focus, several meters before the interaction point, and rotated to kick in the horizontal plane.  When used in combination with the magnetic spectrometer with vertical dispersion, the XTCAV allows for single-shot measurements of the longitudinal phase space. FIG. \ref{fig:XTCAV} shows a simulated XTCAV measurement and the extracted current and momentum profiles for the present XTCAV implementation and the two-bunch beam configuration. The energy difference between the drive and trailing bunches impacts the longitudinal resolution due to the differences in chromatic focusing. However, improved resolution can be achieved for either the trailing or drive bunch individually by setting the spectrometer optics to focus at either bunch energy independently.
With the presently achieved emittance within the experimental area, the longitudinal resolution is limited to approximately \SI{10}{\micro\m}, making it unable to resolve the sub-\SI{}{\micro\m} longitudinal modulations from microbunching on the current profile.

As use of the XTCAV is invasive to the beam incoming to the plasma, it may not be used simultaneously with plasma studies. Non-invasive tools are therefore required to provide single-shot longitudinal information on the incoming beam, such as the separation of the drive and trailing bunches, that may be calibrated with the XTCAV diagnostic.
A machine learning based virtual diagnostic has been developed to predict the shot-to-shot incoming longitudinal phase space and current profiles based on only non-invasive measurements using training data sets acquired with the XTCAV~\cite{Emma_PRAB18}. This diagnostic has been demonstrated at LCLS, and work is underway to implement this technique at FACET-II.

The SYAG beam energy spectrometer and EOS provide direct measurements of the incoming energy and longitudinal profiles that are non-invasive to plasma studies. 
In addition, the EOS system can be configured with a pair of crystals on either side of the beam path to provide time-resolved transverse beam positions on a shot-by-shot basis~\cite{HuntStone_2021}. This system has an ultimate timing resolution of \SI{10}{fs}, and transverse resolution of $<\SI{5}{\micro\m}$ and is capable of resolving the properties of the drive and trailing bunches independently. 

Additional non-intercepting measurements of the incoming bunch length are also acquired on a shot-to-shot basis from a pyroelelectric detector that measures the total intensity of coherent terahertz diffraction radiation after the final bunch compression. The intensity of this radiation can be correlated with relative bunch lengths within the range of 10 to \SI{100}{\micro\m}~\cite{Li_PAC11}.

\subsection{\label{sec:gammas}Betatron Radiation}

As the drive and trailing bunches transit through the plasma cell, electrons in both bunches experience the large transverse forces present within the plasma bubble. This force will induce betatron oscillations in the trajectories of the beam electrons, leading to the emission of betatron radiation in the direction of propagation that is similar to the synchrotron radiation generated in a high-$K$ wiggler~\cite{Corde_RMP13}.  This betatron radiation can be used to diagnose the transverse dynamics of the drive and trailing bunches within the plasma, providing information on matching into the plasma and the transverse hosing instability~\cite{SanMiguel_RSTA18}. For the FACET-II beam parameters, the betatron  radiation is emitted with several milliradian divergence and with a photon energy spectra in the range of keV up to $\sim\SI{1}{MeV}$.

A set of scintillation-based detectors are employed in air in the spectrometer beamline to retrieve angular and spectral information of X-ray and $\gamma$-ray energy photons produced at the IP. Photons with energy below 10-20 keV are stopped by the \SI{5}{mm} aluminum vacuum exit window, but higher energy photons are transmitted through for detection. Their transverse profile is measured by a CCD camera imaging either a uniform scintillator screen  for high resolution measurements, or a CsI pixellated array with $0.5\times0.5$\,mm pixel size ~\footnote{Manufactured by Epic-Crystal} for increased sensitivity.

A second scintillation screen provides spectral information by recording the intensity immediately behind a set of filter materials arranged in a pie shape around the photon-axis. Two of these filter materials act as a pair of Ross filters~\cite{Ross_JOptSocAm28}, with material and thickness chosen for sensitivity to photons of energy $<\SI{100}{keV}$. 
The remaining 10 filters are comprised of various thicknesses of copper up to \SI{8}{mm}, and tungsten up to \SI{3}{mm}, and one segment with no filter material for reference.
The scintillator response behind each filter is impacted by the photon-energy dependent conversion and transmission rates through each material. By determining the intensity behind each filter and comparing to simulated responses using GEANT4 ~\cite{GEANT}, information about the photon energy distribution can be determined, such as the critical energy of a synchrotron-like spectrum. A separate publication summarizes these photon diagnostics and the first results acquired for FACET-II ~\cite{SanMiguel_AAC}.

Additionally, a Compton spectrometer is being developed to provide energy-angular double differential measurements of the betatron radiation in the range of \SI{180}{\kilo\electronvolt} to \SI{28}{\mega\electronvolt}~\cite{Naranjo_IPAC21}. This device will perform the measurement in vacuum, $\sim\SI{2}{\m}$ prior to the spectrometer diagnostics table.

\section{\label{sec:beamtests}Initial Beam Plasma Interaction Studies}

Two decades of beam-plasma interaction experiments have clearly indicated that for a \SI{10}{\giga\electronvolt}-class electron or positron bunch that has nC's of charge, the most sensitive diagnostics of the beam brightness is the plasma itself~\cite{Joshi_SA06}.
A beam with sufficient field intensity can ionize a plasma and generate wakefields in it through interaction with plasma particles.
These interactions change the conditions of both the beam and the plasma first-hand, independent of external diagnostics which have their own limitations such as finite resolution and sensitivity. 
Therefore, the beam-plasma interaction can provide good diagnostics of the incoming beam conditions through changes to the beam spot size, energy loss or gain of different slices of the beam, or the radiation emitted by the charged particles as they traverse a column of gas.

Beam delivery to users at FACET-II started in 2022 for user-assisted commissioning of beam delivery and experimental systems.
Delivery to users was interleaved into the beam commissioning to allow for users to exercise equipment, develop data acquisition techniques, and gain first insights into their experimental programs.
During this phase, the beam was delivered in single bunch mode with nominal bunch charge of \SI{1.6}{\nano\coulomb}, transverse spot sizes down to $\sim$20$\times20$\,\SI{}{\micro\m^2}, and bunch lengths of $\sim$\SI{20}{\micro\m}.

As part of the initial PWFA studies, the single-bunch beam was passed through several meters of gas at the IP to investigate beam ionization and for commissioning of the electron spectrometer and betatron radiation diagnostics.
The differential pumping system was used to maintain a static gas pressure of either hydrogen gas up to \SI{2}{Torr} or helium gas up to \SI{5}{Torr} in the \SI{4}{\m} of beamline between the upstream and downstream beryllium windows.
The beam was focused to a vacuum waist of $\beta=\SI{50}{\centi\m}$ in both x and y planes at a location approximately \SI{0.5}{\m} into the gas column. No laser preionization was employed in the studies presented here.

Despite the measured beam parameters suggesting insufficient beam density for field ionization using ADK theory~\cite{ADK}, we have observed that the present beam conditions are capable of driving a strong wake in both hydrogen and helium gases.
Evidently, the beam diagnostic techniques such as the XTCAV cannot presently resolve the full longitudinal characteristics of the beam.
The field ionization observed in the experiment can be explained by the beam's temporal structure exhibiting one or more strong, short peaks on top of a low current background~\cite{ChaojiePaper}.

\begin{figure}[th]
    \centering
    \includegraphics[width=0.5\textwidth]{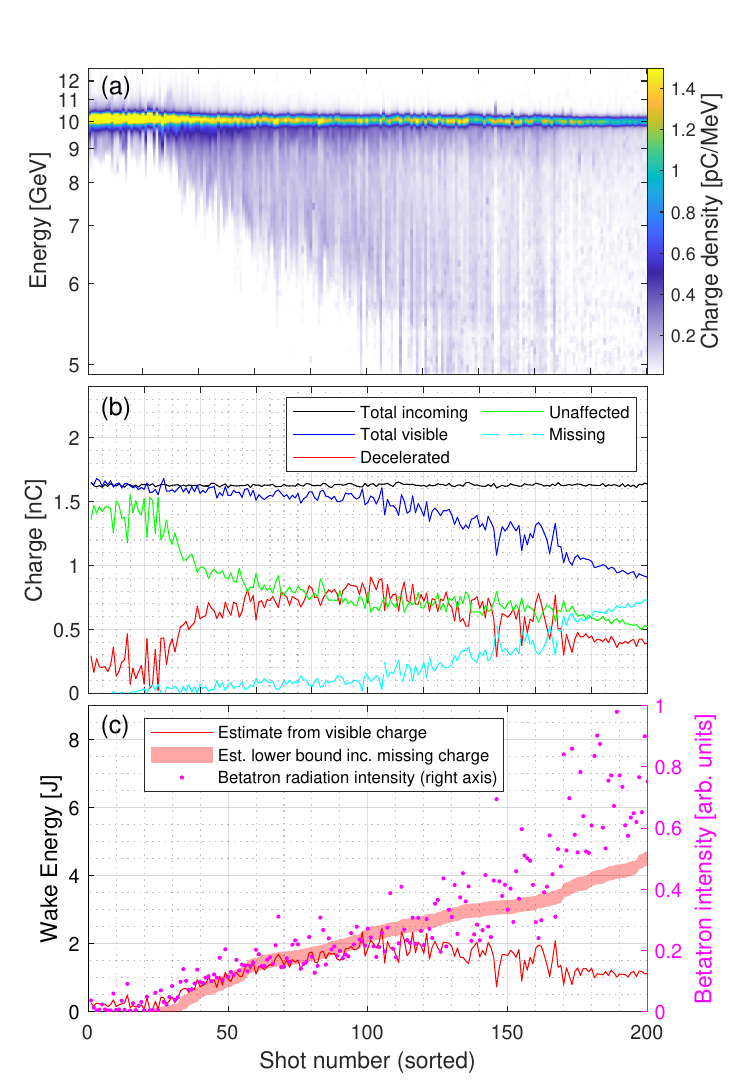}
    \caption{(a) A “waterfall” plot of the electron energy spectra after traversing the plasma. The series of 200 sequential shots acquired at a pressure of \SI{2}{Torr} are sorted by the estimated energy transferred to the plasma wake. 
    (b) Charge breakdown between decelerated and unaffected portions of the spectra.
    For later shots in the series, the electron spectra extend beyond the lower limit of the image, as reflected by the loss of total charge measured. The total incoming charge is measured by a toroidal charge monitor just upstream of the IP.
    (c) The energy estimated to be transferred from the electron beam to the wake, determined from the measured energy spectra. The solid line accounts for only those electrons that are visible on the screen, while the shaded line estimates a lower bound, accounting for missing charge.
    Also overlaid on the plot is the intensity of the betatron radiation measured for each shot, showing strong correlation with wake energy for shots where the total charge is visible.}
    \label{fig:PWFAenergydep}
\end{figure}

The existence of a plasma interaction however is indisputably confirmed by concurrent observations of the plasma emission light measured by cameras viewing the IP, the deceleration of electrons imaged by the electron spectrometer to energies below \SI{2}{\giga\electronvolt}, and the substantial increase in X-ray photons from betatron oscillations within the plasma channel.
With the present beam conditions, the wake intensity was observed to vary substantially with the incoming beam parameters.
FIG. \ref{fig:PWFAenergydep} shows the energy spectra as observed using the large field of view energy spectrometer and the analysis of a series of 200 sequential shots recorded at \SI{10}{Hz} of the beam passing through the \SI{4}{m} hydrogen gas column at \SI{2}{Torr} pressure (plasma density $\sim\SI{6e16}{cm^{-3}}$), sorted by the estimated energy transferred to the wake.

The energy loss of electrons due to the interaction with the wake is shown to vary significantly due to the jitter on the incoming beam parameters. This varies from no significant plasma interaction in several shots, to the deceleration of electrons to below \SI{5}{\giga\electronvolt} - the lower limit of the field of view of the diagnostic with this spectrometer setting.

The single-shot electron energy spectrum enables a distinction between two categories of electrons: those within the beam that remain virtually unaffected by the plasma interaction, residing in the peak centered at \SI{10}{\giga\electronvolt}, and those that undergo deceleration. This categorization is illustrated in FIG. \ref{fig:PWFAenergydep}(b), indicating that  at this pressure, over 50\% of the charge can be actively participating in the plasma interaction.
In shots experiencing the most substantial deceleration, electrons with energies below approximately \SI{5}{\giga\electronvolt} descend beyond the lower boundary of the field of view, resulting in a loss of the total measured charge for these instances.
Another factor contributing to the missing charge is the presence of electrons in less dense regions of the spectra that are not visible above the image background, comprising up to 5\% of the total charge in shots where the spectrum does not extend off the imaging screen.

The energy deposited into the plasma wake is estimated as the difference between the initial energy content of the incoming beam, approximately \SI{16}{J} at \SI{10}{\giga\electronvolt}, and total energy of the electrons after they exit the plasma. 
However, for shots experiencing the greatest energy loss, some electrons fall below the camera's field of view cannot be accounted for directly in this estimate.
We can therefore establish only a lower-bound for the energy transferred to the wake by assuming the missing charge on the spectrometer screen possesses a maximum energy equivalent to the lower cutoff of the field of view, which is \SI{4.9}{\giga\electronvolt}.
By factoring in the missing charge using this approach, we can establish a conservative estimate that suggests at least \SI{5}{J} of energy is transferred to the wake for the shots that experience the largest energy loss.
This corresponds to a minimum effective beam-to-wake transfer efficiency of approximately 50\% from the \SI{1}{\nano\coulomb} of charge that participates in the interaction.

The intensity of betatron radiation is superimposed on the same plot as the estimated wake energy, revealing a strong correlation with the energy transferred to the wake in shots where the majority of the charge is visible on the screen.
While there is not a conclusive argument that this correlation should always hold true, this correlation extends to follow the estimated lower limit for shots that experience significant energy loss. This trend suggests that the energy transferred to the wake is indeed higher than the lower bound estimated by visible charge alone.

\begin{figure}[th]
    \centering
    \includegraphics[width=0.5\textwidth]{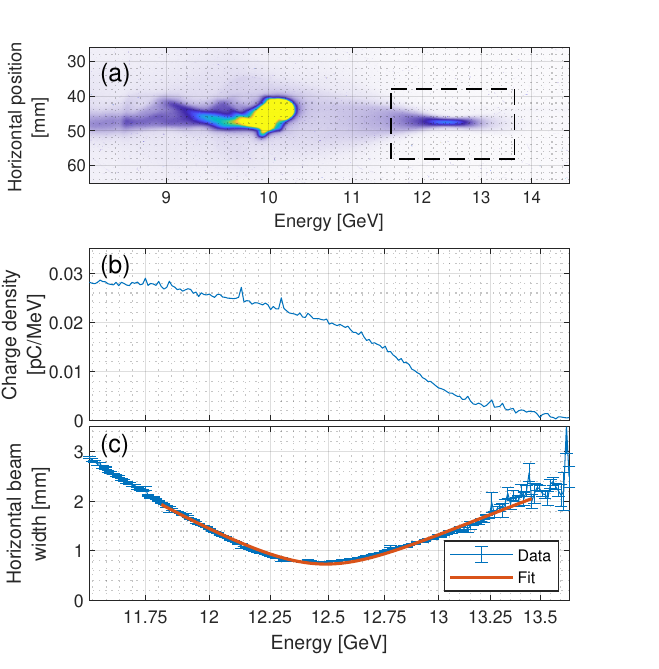}
    \caption{
    (a) The electron energy spectrum with the spectrometer set to reimage at an energy of \SI{12.5}{\giga\electronvolt}. Energy depleted electrons are visible at energies below \SI{10}{\giga\electronvolt}, while some 10s of pC of charge are accelerated up to $\sim\SI{13.5}{\giga\electronvolt}$ in this shot.
    An emittance analysis was performed for the charge indicated by the box, with charge distribution shown in (b).
    (c) The beam width as a function of energy, and the emittance fit function overlaid which provides a normalized emittance of approximately \SI{1500}{\micro\m}. The Twiss parameters determined from the fit indicate that the beam waist (and hence the exit from the plasma) was located at the location of the Beryllium window.}
    \label{fig:BF}
\end{figure}

Evidence of acceleration by PWFA was measured by imaging the electron spectra at energies above \SI{10}{\giga\electronvolt} using the large field of view electron spectrometer. FIG. \ref{fig:BF} shows the electron spectrum acquired with the spectrometer quadrupoles set to image \SI{12.5}{\giga\electronvolt} electrons from the end of the gas column. This spectrum shows decelerated electrons with energy $<\SI{10}{\giga\electronvolt}$, indicating the presence of a strong wake generation, and charge extending to beyond \SI{13}{\giga\electronvolt} in this shot. We infer that this accelerated charge originated from the small fraction of electrons far within the tail of the single bunch that experiences the accelerating phase of the plasma wakefields.

Performing a single-shot emittance measurement on the charge around \SI{12.5}{\giga\electronvolt}, as described in the prior section, provides both a normalized emittance value of approximately \SI{1500}{\micro\m}, and places the exit from the plasma precisely at the location of the end of the gas column, with a waist $\beta$ of \SI{17}{cm}.
This measurement is not resolution limited, as the imaging resolution of this profile monitor is $<\SI{100}{\micro\meter}$, far below the minimum spot size.
The emittance measured here is extremely large, in part due to the presumed large emittance of the electrons far within the tail of the bunch, and also due to no effort made in matching these electrons to the plasma.
This analysis however serves as a first implementation of this single-shot emittance measurement in the new FACET-II beamline, and provides useful information about the length of the plasma.
The overall length of the plasma can be determined to extend at least \SI{3}{m} from the first IP camera that detects plasma light near the incoming beam waist, to the location of the measured waist position at the plasma exit.

\section{Conclusion}

We have described the experimental setup, accelerator parameters, and beam diagnostics required to demonstrate a single stage plasma wakefield accelerator at FACET-II, which will approach the parameters required for linear colliders and high brightness light sources.
In our preliminary investigations of the beam-plasma interaction with single bunches, we have observed the formation of beam-ionized plasmas extending for several meters in length within hydrogen gas.
Measurements of the resulting electron spectrum, plasma emission light, and betatron radiation indicate significant energy transfer from the drive beam to the wake with the present beam conditions.

Ongoing efforts focus on achieving efficient and stable energy transfer from the drive beam to the wake within both lithium and hydrogen plasmas as the electron beam conditions continue to improve with further beam development.
Once the two-bunch configuration is operational, the ultimate goal is to demonstrate multi-GeV energy doubling of the trailing bunch, preserving both emittance and energy spread as required for accelerator applications.

\begin{acknowledgments}

FACET-II is supported in part by the U.S. Department of Energy under contract number DE-AC02-76SF00515.
This work is supported by the U.S. Department of Energy grant DE-SC0010064:0011 and U.S. National Foundation grant 2108970 at UCLA.
E. Gerstmayr was supported by the U.S. Department of Energy, Office of Science, Fusion Energy Sciences under Award DE-SC0020076.
A.\,Knetsch and P. San Miguel Claveria were supported by the France-Stanford Center for Interdisciplinary Studies for their travels to SLAC National Accelerator Laboratory.
H. Ekerfelt was supported by the Knut and Alice Wallenberg Foundation (KAW 2018.0450).

\end{acknowledgments}

%% file: main.bbl
\begin{thebibliography}{50}%
\makeatletter
\providecommand \@ifxundefined [1]{%
 \@ifx{#1\undefined}
}%
\providecommand \@ifnum [1]{%
 \ifnum #1\expandafter \@firstoftwo
 \else \expandafter \@secondoftwo
 \fi
}%
\providecommand \@ifx [1]{%
 \ifx #1\expandafter \@firstoftwo
 \else \expandafter \@secondoftwo
 \fi
}%
\providecommand \natexlab [1]{#1}%
\providecommand \enquote  [1]{``#1''}%
\providecommand \bibnamefont  [1]{#1}%
\providecommand \bibfnamefont [1]{#1}%
\providecommand \citenamefont [1]{#1}%
\providecommand \href@noop [0]{\@secondoftwo}%
\providecommand \href [0]{\begingroup \@sanitize@url \@href}%
\providecommand \@href[1]{\@@startlink{#1}\@@href}%
\providecommand \@@href[1]{\endgroup#1\@@endlink}%
\providecommand \@sanitize@url [0]{\catcode `\\12\catcode `\$12\catcode
  `\&12\catcode `\#12\catcode `\^12\catcode `\_12\catcode `\%12\relax}%
\providecommand \@@startlink[1]{}%
\providecommand \@@endlink[0]{}%
\providecommand \url  [0]{\begingroup\@sanitize@url \@url }%
\providecommand \@url [1]{\endgroup\@href {#1}{\urlprefix }}%
\providecommand \urlprefix  [0]{URL }%
\providecommand \Eprint [0]{\href }%
\providecommand \doibase [0]{https://doi.org/}%
\providecommand \selectlanguage [0]{\@gobble}%
\providecommand \bibinfo  [0]{\@secondoftwo}%
\providecommand \bibfield  [0]{\@secondoftwo}%
\providecommand \translation [1]{[#1]}%
\providecommand \BibitemOpen [0]{}%
\providecommand \bibitemStop [0]{}%
\providecommand \bibitemNoStop [0]{.\EOS\space}%
\providecommand \EOS [0]{\spacefactor3000\relax}%
\providecommand \BibitemShut  [1]{\csname bibitem#1\endcsname}%
\let\auto@bib@innerbib\@empty
\bibitem [{\citenamefont {Adolphsen}\ \emph {et~al.}(2013)\citenamefont
  {Adolphsen} \emph {et~al.}}]{ILC_TDR}%
  \BibitemOpen
  \bibfield  {author} {\bibinfo {author} {\bibfnamefont {C.}~\bibnamefont
  {Adolphsen}} \emph {et~al.},\ }\href@noop {} {\emph {\bibinfo {title} {{The
  International Linear Collider Technical Design Report}}}},\ \bibinfo {type}
  {Tech. Rep.}\ (\bibinfo {year} {2013})\ \Eprint
  {https://arxiv.org/abs/1306.6328} {arXiv:1306.6328} \BibitemShut {NoStop}%
\bibitem [{\citenamefont {Dasu}\ \emph {et~al.}(2022)\citenamefont {Dasu} \emph
  {et~al.}}]{CCC}%
  \BibitemOpen
  \bibfield  {author} {\bibinfo {author} {\bibfnamefont {S.}~\bibnamefont
  {Dasu}} \emph {et~al.},\ }\bibfield  {title} {\bibinfo {title} {{Strategy for
  Understanding the Higgs Physics: The Cool Copper Collider}},\ }in\ \href@noop
  {} {\emph {\bibinfo {booktitle} {{Proceedings of the US Community Study on
  the Future of Particle Physics (Snowmass 2021)}}}}\ (\bibinfo {year} {2022})\
  \Eprint {https://arxiv.org/abs/2203.07646} {arXiv:2203.07646 [hep-ex]}
  \BibitemShut {NoStop}%
\bibitem [{\citenamefont {Charles}\ \emph {et~al.}(2018)\citenamefont {Charles}
  \emph {et~al.}}]{CLIC}%
  \BibitemOpen
  \bibfield  {author} {\bibinfo {author} {\bibfnamefont {T.~K.}\ \bibnamefont
  {Charles}} \emph {et~al.} (\bibinfo {collaboration} {CLICdp, CLIC}),\
  }\bibfield  {title} {\bibinfo {title} {{The Compact Linear Collider (CLIC) -
  2018 Summary Report}}\ }\href {https://doi.org/10.23731/CYRM-2018-002}
  {10.23731/CYRM-2018-002} (\bibinfo {year} {2018})\BibitemShut {NoStop}%
\bibitem [{\citenamefont {Narain}\ \emph {et~al.}(2022)\citenamefont {Narain}
  \emph {et~al.}}]{SMstudies}%
  \BibitemOpen
  \bibfield  {author} {\bibinfo {author} {\bibfnamefont {M.}~\bibnamefont
  {Narain}} \emph {et~al.},\ }\bibfield  {title} {\bibinfo {title} {{The Future
  of US Particle Physics - The Snowmass 2021 Energy Frontier Report}},\
  }\href@noop {} {\  (\bibinfo {year} {2022})},\ \Eprint
  {https://arxiv.org/abs/2211.11084} {arXiv:2211.11084 [hep-ex]} \BibitemShut
  {NoStop}%
\bibitem [{\citenamefont {Chen}\ \emph {et~al.}(1985)\citenamefont {Chen},
  \citenamefont {Dawson}, \citenamefont {Huff},\ and\ \citenamefont
  {Katsouleas}}]{Chen_PRL85}%
  \BibitemOpen
  \bibfield  {author} {\bibinfo {author} {\bibfnamefont {P.}~\bibnamefont
  {Chen}}, \bibinfo {author} {\bibfnamefont {J.~M.}\ \bibnamefont {Dawson}},
  \bibinfo {author} {\bibfnamefont {R.~W.}\ \bibnamefont {Huff}},\ and\
  \bibinfo {author} {\bibfnamefont {T.}~\bibnamefont {Katsouleas}},\ }\bibfield
   {title} {\bibinfo {title} {Acceleration of electrons by the interaction of a
  bunched electron beam with a plasma},\ }\href
  {https://doi.org/10.1103/PhysRevLett.54.693} {\bibfield  {journal} {\bibinfo
  {journal} {Phys. Rev. Lett.}\ }\textbf {\bibinfo {volume} {54}},\ \bibinfo
  {pages} {693} (\bibinfo {year} {1985})}\BibitemShut {NoStop}%
\bibitem [{\citenamefont {Tajima}\ and\ \citenamefont
  {Dawson}(1979)}]{Tajima_PRL79}%
  \BibitemOpen
  \bibfield  {author} {\bibinfo {author} {\bibfnamefont {T.}~\bibnamefont
  {Tajima}}\ and\ \bibinfo {author} {\bibfnamefont {J.~M.}\ \bibnamefont
  {Dawson}},\ }\bibfield  {title} {\bibinfo {title} {Laser electron
  accelerator},\ }\href {https://doi.org/10.1103/PhysRevLett.43.267} {\bibfield
   {journal} {\bibinfo  {journal} {Phys. Rev. Lett.}\ }\textbf {\bibinfo
  {volume} {43}},\ \bibinfo {pages} {267} (\bibinfo {year} {1979})}\BibitemShut
  {NoStop}%
\bibitem [{\citenamefont {Blumenfeld}\ \emph {et~al.}(2007)\citenamefont
  {Blumenfeld} \emph {et~al.}}]{Blumenfeld_Nature07}%
  \BibitemOpen
  \bibfield  {author} {\bibinfo {author} {\bibfnamefont {I.}~\bibnamefont
  {Blumenfeld}} \emph {et~al.},\ }\bibfield  {title} {\bibinfo {title} {{Energy
  doubling of 42 GeV electrons in a metre-scale plasma wakefield
  accelerator}},\ }\href {https://doi.org/10.1038/nature05538} {\bibfield
  {journal} {\bibinfo  {journal} {Nature}\ }\textbf {\bibinfo {volume} {445}},\
  \bibinfo {pages} {741} (\bibinfo {year} {2007})}\BibitemShut {NoStop}%
\bibitem [{\citenamefont {Litos}\ \emph {et~al.}(2016)\citenamefont {Litos}
  \emph {et~al.}}]{Litos_PPCF2016}%
  \BibitemOpen
  \bibfield  {author} {\bibinfo {author} {\bibfnamefont {M.}~\bibnamefont
  {Litos}} \emph {et~al.},\ }\bibfield  {title} {\bibinfo {title} {{9 GeV
  energy gain in a beam-driven plasma wakefield accelerator}},\ }\href
  {https://doi.org/10.1088/0741-3335/58/3/034017} {\bibfield  {journal}
  {\bibinfo  {journal} {Plasma Physics and Controlled Fusion}\ }\textbf
  {\bibinfo {volume} {58}},\ \bibinfo {pages} {034017} (\bibinfo {year}
  {2016})}\BibitemShut {NoStop}%
\bibitem [{\citenamefont {Lindstr\o{}m}\ \emph {et~al.}(2021)\citenamefont
  {Lindstr\o{}m} \emph {et~al.}}]{Lindstrom_PRL21}%
  \BibitemOpen
  \bibfield  {author} {\bibinfo {author} {\bibfnamefont {C.~A.}\ \bibnamefont
  {Lindstr\o{}m}} \emph {et~al.},\ }\bibfield  {title} {\bibinfo {title}
  {Energy-spread preservation and high efficiency in a plasma-wakefield
  accelerator},\ }\href {https://doi.org/10.1103/PhysRevLett.126.014801}
  {\bibfield  {journal} {\bibinfo  {journal} {Phys. Rev. Lett.}\ }\textbf
  {\bibinfo {volume} {126}},\ \bibinfo {pages} {014801} (\bibinfo {year}
  {2021})}\BibitemShut {NoStop}%
\bibitem [{\citenamefont {Lindstrøm}\ \emph {et~al.}(2022)\citenamefont
  {Lindstrøm} \emph {et~al.}}]{Lindstrom_Emit22}%
  \BibitemOpen
  \bibfield  {author} {\bibinfo {author} {\bibfnamefont {C.~A.}\ \bibnamefont
  {Lindstrøm}} \emph {et~al.},\ }\href
  {https://doi.org/10.21203/rs.3.rs-2300900/v1} {\bibinfo {title} {Preservation
  of beam quality in a plasma-wakefield accelerator}} (\bibinfo {year}
  {2022}),\ \Eprint
  {https://arxiv.org/abs/https://www.researchsquare.com/article/rs-2300900/v1}
  {https://www.researchsquare.com/article/rs-2300900/v1} \BibitemShut {NoStop}%
\bibitem [{\citenamefont {Yakimenko}\ \emph {et~al.}(2019)\citenamefont
  {Yakimenko} \emph {et~al.}}]{Yakimenko_PRAB19}%
  \BibitemOpen
  \bibfield  {author} {\bibinfo {author} {\bibfnamefont {V.}~\bibnamefont
  {Yakimenko}} \emph {et~al.},\ }\bibfield  {title} {\bibinfo {title}
  {{FACET-II facility for advanced accelerator experimental tests}},\ }\href
  {https://doi.org/10.1103/PhysRevAccelBeams.22.101301} {\bibfield  {journal}
  {\bibinfo  {journal} {Phys. Rev. Accel. Beams}\ }\textbf {\bibinfo {volume}
  {22}},\ \bibinfo {pages} {101301} (\bibinfo {year} {2019})}\BibitemShut
  {NoStop}%
\bibitem [{\citenamefont {Joshi}\ \emph {et~al.}(2018)\citenamefont {Joshi}
  \emph {et~al.}}]{Joshi_2018}%
  \BibitemOpen
  \bibfield  {author} {\bibinfo {author} {\bibfnamefont {C.}~\bibnamefont
  {Joshi}} \emph {et~al.},\ }\bibfield  {title} {\bibinfo {title} {Plasma
  wakefield acceleration experiments at {FACET} {II}},\ }\href
  {https://doi.org/10.1088/1361-6587/aaa2e3} {\bibfield  {journal} {\bibinfo
  {journal} {Plasma Physics and Controlled Fusion}\ }\textbf {\bibinfo {volume}
  {60}},\ \bibinfo {pages} {034001} (\bibinfo {year} {2018})}\BibitemShut
  {NoStop}%
\bibitem [{\citenamefont {Pe\~na}\ \emph {et~al.}(2023)\citenamefont {Pe\~na}
  \emph {et~al.}}]{Pena_23}%
  \BibitemOpen
  \bibfield  {author} {\bibinfo {author} {\bibfnamefont {F.}~\bibnamefont
  {Pe\~na}} \emph {et~al.},\ }\bibfield  {title} {\bibinfo {title} {{Energy
  Depletion and Re-Acceleration of Driver Electrons in a Plasma-Wakefield
  Accelerator}},\ }\href@noop {} {\  (\bibinfo {year} {2023})},\ \Eprint
  {https://arxiv.org/abs/2305.09581} {arXiv:2305.09581 [physics.acc-ph]}
  \BibitemShut {NoStop}%
\bibitem [{\citenamefont {Katsouleas}\ \emph {et~al.}(1987)\citenamefont
  {Katsouleas}, \citenamefont {Wilks}, \citenamefont {Chen}, \citenamefont
  {Dawson},\ and\ \citenamefont {Su}}]{Katsouleas_PA87}%
  \BibitemOpen
  \bibfield  {author} {\bibinfo {author} {\bibfnamefont {T.~C.}\ \bibnamefont
  {Katsouleas}}, \bibinfo {author} {\bibfnamefont {S.}~\bibnamefont {Wilks}},
  \bibinfo {author} {\bibfnamefont {P.}~\bibnamefont {Chen}}, \bibinfo {author}
  {\bibfnamefont {J.~M.}\ \bibnamefont {Dawson}},\ and\ \bibinfo {author}
  {\bibfnamefont {J.~J.}\ \bibnamefont {Su}},\ }\bibfield  {title} {\bibinfo
  {title} {{Beam Loading in Plasma Accelerators}},\ }\href@noop {} {\bibfield
  {journal} {\bibinfo  {journal} {Part. Accel.}\ }\textbf {\bibinfo {volume}
  {22}},\ \bibinfo {pages} {81} (\bibinfo {year} {1987})}\BibitemShut {NoStop}%
\bibitem [{\citenamefont {Chen}\ \emph {et~al.}(1986)\citenamefont {Chen},
  \citenamefont {Su}, \citenamefont {Dawson}, \citenamefont {Bane},\ and\
  \citenamefont {Wilson}}]{Chen_PRL86}%
  \BibitemOpen
  \bibfield  {author} {\bibinfo {author} {\bibfnamefont {P.}~\bibnamefont
  {Chen}}, \bibinfo {author} {\bibfnamefont {J.~J.}\ \bibnamefont {Su}},
  \bibinfo {author} {\bibfnamefont {J.~M.}\ \bibnamefont {Dawson}}, \bibinfo
  {author} {\bibfnamefont {K.~L.~F.}\ \bibnamefont {Bane}},\ and\ \bibinfo
  {author} {\bibfnamefont {P.~B.}\ \bibnamefont {Wilson}},\ }\bibfield  {title}
  {\bibinfo {title} {Energy transfer in the plasma wake-field accelerator},\
  }\href {https://doi.org/10.1103/PhysRevLett.56.1252} {\bibfield  {journal}
  {\bibinfo  {journal} {Phys. Rev. Lett.}\ }\textbf {\bibinfo {volume} {56}},\
  \bibinfo {pages} {1252} (\bibinfo {year} {1986})}\BibitemShut {NoStop}%
\bibitem [{\citenamefont {Lotov}(2005)}]{Lotov_PoP05}%
  \BibitemOpen
  \bibfield  {author} {\bibinfo {author} {\bibfnamefont {K.~V.}\ \bibnamefont
  {Lotov}},\ }\bibfield  {title} {\bibinfo {title} {{Efficient operating mode
  of the plasma wakefield accelerator}},\ }\bibfield  {journal} {\bibinfo
  {journal} {Physics of Plasmas}\ }\textbf {\bibinfo {volume} {12}},\ \href
  {https://doi.org/10.1063/1.1889444} {10.1063/1.1889444} (\bibinfo {year}
  {2005}),\ \bibinfo {note} {053105}\BibitemShut {NoStop}%
\bibitem [{\citenamefont {Tzoufras}\ \emph {et~al.}(2008)\citenamefont
  {Tzoufras} \emph {et~al.}}]{Tzoufras_PRL08}%
  \BibitemOpen
  \bibfield  {author} {\bibinfo {author} {\bibfnamefont {M.}~\bibnamefont
  {Tzoufras}} \emph {et~al.},\ }\bibfield  {title} {\bibinfo {title} {Beam
  loading in the nonlinear regime of plasma-based acceleration},\ }\href
  {https://doi.org/10.1103/PhysRevLett.101.145002} {\bibfield  {journal}
  {\bibinfo  {journal} {Phys. Rev. Lett.}\ }\textbf {\bibinfo {volume} {101}},\
  \bibinfo {pages} {145002} (\bibinfo {year} {2008})}\BibitemShut {NoStop}%
\bibitem [{\citenamefont {Xu}\ \emph {et~al.}(2016)\citenamefont {Xu} \emph
  {et~al.}}]{Xu_RPL16}%
  \BibitemOpen
  \bibfield  {author} {\bibinfo {author} {\bibfnamefont {X.~L.}\ \bibnamefont
  {Xu}} \emph {et~al.},\ }\bibfield  {title} {\bibinfo {title} {Physics of
  phase space matching for staging plasma and traditional accelerator
  components using longitudinally tailored plasma profiles},\ }\href
  {https://doi.org/10.1103/PhysRevLett.116.124801} {\bibfield  {journal}
  {\bibinfo  {journal} {Phys. Rev. Lett.}\ }\textbf {\bibinfo {volume} {116}},\
  \bibinfo {pages} {124801} (\bibinfo {year} {2016})}\BibitemShut {NoStop}%
\bibitem [{\citenamefont {Th\'evenet}\ \emph {et~al.}(2019)\citenamefont
  {Th\'evenet}, \citenamefont {Lehe}, \citenamefont {Schroeder}, \citenamefont
  {Benedetti}, \citenamefont {Vay}, \citenamefont {Esarey},\ and\ \citenamefont
  {Leemans}}]{Thevenet_PRAB19}%
  \BibitemOpen
  \bibfield  {author} {\bibinfo {author} {\bibfnamefont {M.}~\bibnamefont
  {Th\'evenet}}, \bibinfo {author} {\bibfnamefont {R.}~\bibnamefont {Lehe}},
  \bibinfo {author} {\bibfnamefont {C.~B.}\ \bibnamefont {Schroeder}}, \bibinfo
  {author} {\bibfnamefont {C.}~\bibnamefont {Benedetti}}, \bibinfo {author}
  {\bibfnamefont {J.-L.}\ \bibnamefont {Vay}}, \bibinfo {author} {\bibfnamefont
  {E.}~\bibnamefont {Esarey}},\ and\ \bibinfo {author} {\bibfnamefont {W.~P.}\
  \bibnamefont {Leemans}},\ }\bibfield  {title} {\bibinfo {title} {Emittance
  growth due to misalignment in multistage laser-plasma accelerators},\ }\href
  {https://doi.org/10.1103/PhysRevAccelBeams.22.051302} {\bibfield  {journal}
  {\bibinfo  {journal} {Phys. Rev. Accel. Beams}\ }\textbf {\bibinfo {volume}
  {22}},\ \bibinfo {pages} {051302} (\bibinfo {year} {2019})}\BibitemShut
  {NoStop}%
\bibitem [{\citenamefont {Ariniello}\ \emph {et~al.}(2019)\citenamefont
  {Ariniello}, \citenamefont {Doss}, \citenamefont {Hunt-Stone}, \citenamefont
  {Cary},\ and\ \citenamefont {Litos}}]{Ariniello_PRAB19}%
  \BibitemOpen
  \bibfield  {author} {\bibinfo {author} {\bibfnamefont {R.}~\bibnamefont
  {Ariniello}}, \bibinfo {author} {\bibfnamefont {C.~E.}\ \bibnamefont {Doss}},
  \bibinfo {author} {\bibfnamefont {K.}~\bibnamefont {Hunt-Stone}}, \bibinfo
  {author} {\bibfnamefont {J.~R.}\ \bibnamefont {Cary}},\ and\ \bibinfo
  {author} {\bibfnamefont {M.~D.}\ \bibnamefont {Litos}},\ }\bibfield  {title}
  {\bibinfo {title} {Transverse beam dynamics in a plasma density ramp},\
  }\href {https://doi.org/10.1103/PhysRevAccelBeams.22.041304} {\bibfield
  {journal} {\bibinfo  {journal} {Phys. Rev. Accel. Beams}\ }\textbf {\bibinfo
  {volume} {22}},\ \bibinfo {pages} {041304} (\bibinfo {year}
  {2019})}\BibitemShut {NoStop}%
\bibitem [{\citenamefont {Ariniello}\ \emph {et~al.}(2022)\citenamefont
  {Ariniello}, \citenamefont {Doss}, \citenamefont {Lee}, \citenamefont
  {Hansel}, \citenamefont {Cary},\ and\ \citenamefont
  {Litos}}]{Ariniello_PRR22}%
  \BibitemOpen
  \bibfield  {author} {\bibinfo {author} {\bibfnamefont {R.}~\bibnamefont
  {Ariniello}}, \bibinfo {author} {\bibfnamefont {C.~E.}\ \bibnamefont {Doss}},
  \bibinfo {author} {\bibfnamefont {V.}~\bibnamefont {Lee}}, \bibinfo {author}
  {\bibfnamefont {C.}~\bibnamefont {Hansel}}, \bibinfo {author} {\bibfnamefont
  {J.~R.}\ \bibnamefont {Cary}},\ and\ \bibinfo {author} {\bibfnamefont
  {M.~D.}\ \bibnamefont {Litos}},\ }\bibfield  {title} {\bibinfo {title}
  {Chromatic transverse dynamics in a nonlinear plasma accelerator},\ }\href
  {https://doi.org/10.1103/PhysRevResearch.4.043120} {\bibfield  {journal}
  {\bibinfo  {journal} {Phys. Rev. Res.}\ }\textbf {\bibinfo {volume} {4}},\
  \bibinfo {pages} {043120} (\bibinfo {year} {2022})}\BibitemShut {NoStop}%
\bibitem [{\citenamefont {An}\ \emph {et~al.}(2017)\citenamefont {An},
  \citenamefont {Lu}, \citenamefont {Huang}, \citenamefont {Xu}, \citenamefont
  {Hogan}, \citenamefont {Joshi},\ and\ \citenamefont {Mori}}]{An_PRL17}%
  \BibitemOpen
  \bibfield  {author} {\bibinfo {author} {\bibfnamefont {W.}~\bibnamefont
  {An}}, \bibinfo {author} {\bibfnamefont {W.}~\bibnamefont {Lu}}, \bibinfo
  {author} {\bibfnamefont {C.}~\bibnamefont {Huang}}, \bibinfo {author}
  {\bibfnamefont {X.}~\bibnamefont {Xu}}, \bibinfo {author} {\bibfnamefont
  {M.~J.}\ \bibnamefont {Hogan}}, \bibinfo {author} {\bibfnamefont
  {C.}~\bibnamefont {Joshi}},\ and\ \bibinfo {author} {\bibfnamefont {W.~B.}\
  \bibnamefont {Mori}},\ }\bibfield  {title} {\bibinfo {title} {Ion motion
  induced emittance growth of matched electron beams in plasma wakefields},\
  }\href {https://doi.org/10.1103/PhysRevLett.118.244801} {\bibfield  {journal}
  {\bibinfo  {journal} {Phys. Rev. Lett.}\ }\textbf {\bibinfo {volume} {118}},\
  \bibinfo {pages} {244801} (\bibinfo {year} {2017})}\BibitemShut {NoStop}%
\bibitem [{\citenamefont {Zhao}\ \emph {et~al.}(2020)\citenamefont {Zhao},
  \citenamefont {Lehe}, \citenamefont {Myers}, \citenamefont {Thévenet},
  \citenamefont {Huebl}, \citenamefont {Schroeder},\ and\ \citenamefont
  {Vay}}]{Zhao_PoP20}%
  \BibitemOpen
  \bibfield  {author} {\bibinfo {author} {\bibfnamefont {Y.}~\bibnamefont
  {Zhao}}, \bibinfo {author} {\bibfnamefont {R.}~\bibnamefont {Lehe}}, \bibinfo
  {author} {\bibfnamefont {A.}~\bibnamefont {Myers}}, \bibinfo {author}
  {\bibfnamefont {M.}~\bibnamefont {Thévenet}}, \bibinfo {author}
  {\bibfnamefont {A.}~\bibnamefont {Huebl}}, \bibinfo {author} {\bibfnamefont
  {C.~B.}\ \bibnamefont {Schroeder}},\ and\ \bibinfo {author} {\bibfnamefont
  {J.-L.}\ \bibnamefont {Vay}},\ }\bibfield  {title} {\bibinfo {title}
  {{Modeling of emittance growth due to Coulomb collisions in plasma-based
  accelerators}},\ }\bibfield  {journal} {\bibinfo  {journal} {Physics of
  Plasmas}\ }\textbf {\bibinfo {volume} {27}},\ \href
  {https://doi.org/10.1063/5.0023776} {10.1063/5.0023776} (\bibinfo {year}
  {2020}),\ \bibinfo {note} {113105}\BibitemShut {NoStop}%
\bibitem [{\citenamefont {Li}\ \emph {et~al.}(2021)\citenamefont {Li},
  \citenamefont {An}, \citenamefont {Decyk}, \citenamefont {Xu}, \citenamefont
  {Hogan},\ and\ \citenamefont {Mori}}]{QPAD}%
  \BibitemOpen
  \bibfield  {author} {\bibinfo {author} {\bibfnamefont {F.}~\bibnamefont
  {Li}}, \bibinfo {author} {\bibfnamefont {W.}~\bibnamefont {An}}, \bibinfo
  {author} {\bibfnamefont {V.~K.}\ \bibnamefont {Decyk}}, \bibinfo {author}
  {\bibfnamefont {X.}~\bibnamefont {Xu}}, \bibinfo {author} {\bibfnamefont
  {M.~J.}\ \bibnamefont {Hogan}},\ and\ \bibinfo {author} {\bibfnamefont
  {W.~B.}\ \bibnamefont {Mori}},\ }\bibfield  {title} {\bibinfo {title} {A
  quasi-static particle-in-cell algorithm based on an azimuthal fourier
  decomposition for highly efficient simulations of plasma-based acceleration:
  Qpad},\ }\href {https://doi.org/https://doi.org/10.1016/j.cpc.2020.107784}
  {\bibfield  {journal} {\bibinfo  {journal} {Computer Physics Communications}\
  }\textbf {\bibinfo {volume} {261}},\ \bibinfo {pages} {107784} (\bibinfo
  {year} {2021})}\BibitemShut {NoStop}%
\bibitem [{\citenamefont {{Ammosov}}\ \emph {et~al.}(1986)\citenamefont
  {{Ammosov}}, \citenamefont {{Delone}},\ and\ \citenamefont
  {{Krainov}}}]{ADK}%
  \BibitemOpen
  \bibfield  {author} {\bibinfo {author} {\bibfnamefont {M.~V.}\ \bibnamefont
  {{Ammosov}}}, \bibinfo {author} {\bibfnamefont {N.~B.}\ \bibnamefont
  {{Delone}}},\ and\ \bibinfo {author} {\bibfnamefont {V.~P.}\ \bibnamefont
  {{Krainov}}},\ }\bibfield  {title} {\bibinfo {title} {{Tunnel ionization of
  complex atoms and of atomic ions in an alternating electromagnetic field}},\
  }\href@noop {} {\bibfield  {journal} {\bibinfo  {journal} {Soviet Journal of
  Experimental and Theoretical Physics}\ }\textbf {\bibinfo {volume} {64}},\
  \bibinfo {pages} {1191} (\bibinfo {year} {1986})}\BibitemShut {NoStop}%
\bibitem [{FAC(2016)}]{FACET_TDR}%
  \BibitemOpen
  \href {https://doi.org/10.2172/1340171} {\emph {\bibinfo {title} {{Technical
  Design Report for the FACET-II Project at SLAC National Accelerator
  Laboratory}}}},\ \bibinfo {type} {Tech. Rep.}\ (\bibinfo  {institution}
  {{SLAC National Accelerator Laboratory}},\ \bibinfo {year}
  {2016})\BibitemShut {NoStop}%
\bibitem [{\citenamefont {Huang}\ \emph {et~al.}(2010)\citenamefont {Huang}
  \emph {et~al.}}]{Huang_PRAB10}%
  \BibitemOpen
  \bibfield  {author} {\bibinfo {author} {\bibfnamefont {Z.}~\bibnamefont
  {Huang}} \emph {et~al.},\ }\bibfield  {title} {\bibinfo {title} {Measurements
  of the linac coherent light source laser heater and its impact on the x-ray
  free-electron laser performance},\ }\href
  {https://doi.org/10.1103/PhysRevSTAB.13.020703} {\bibfield  {journal}
  {\bibinfo  {journal} {Phys. Rev. ST Accel. Beams}\ }\textbf {\bibinfo
  {volume} {13}},\ \bibinfo {pages} {020703} (\bibinfo {year}
  {2010})}\BibitemShut {NoStop}%
\bibitem [{\citenamefont {Storey}\ \emph {et~al.}(2023)\citenamefont {Storey}
  \emph {et~al.}}]{Storey_IPAC23}%
  \BibitemOpen
  \bibfield  {author} {\bibinfo {author} {\bibfnamefont {D.}~\bibnamefont
  {Storey}} \emph {et~al.},\ }\bibfield  {title} {\bibinfo {title} {{Status and
  first results from FACET-II towards the demonstration of plasma wakefield
  acceleration, coherent radiation generation, and probing strong-field QED}},\
  }in\ \href@noop {} {\emph {\bibinfo {booktitle} {Proceedings of the 14th
  International Particle Accelerator Conference}}}\ (\bibinfo {year}
  {2023})\BibitemShut {NoStop}%
\bibitem [{\citenamefont {Vafaei-Najafabadi}\ \emph {et~al.}(2012)\citenamefont
  {Vafaei-Najafabadi}, \citenamefont {Shaw}, \citenamefont {Marsh},
  \citenamefont {Joshi},\ and\ \citenamefont {Hogan}}]{Navid_AAC12}%
  \BibitemOpen
  \bibfield  {author} {\bibinfo {author} {\bibfnamefont {N.}~\bibnamefont
  {Vafaei-Najafabadi}}, \bibinfo {author} {\bibfnamefont {J.~L.}\ \bibnamefont
  {Shaw}}, \bibinfo {author} {\bibfnamefont {K.~A.}\ \bibnamefont {Marsh}},
  \bibinfo {author} {\bibfnamefont {C.}~\bibnamefont {Joshi}},\ and\ \bibinfo
  {author} {\bibfnamefont {M.~J.}\ \bibnamefont {Hogan}},\ }\bibfield  {title}
  {\bibinfo {title} {{Meter scale plasma source for plasma wakefield
  experiments}},\ }\href {https://doi.org/10.1063/1.4773774} {\bibfield
  {journal} {\bibinfo  {journal} {AIP Conference Proceedings}\ }\textbf
  {\bibinfo {volume} {1507}},\ \bibinfo {pages} {650} (\bibinfo {year}
  {2012})}\BibitemShut {NoStop}%
\bibitem [{\citenamefont {Vafaei-Najafabadi}\ \emph {et~al.}(2014)\citenamefont
  {Vafaei-Najafabadi} \emph {et~al.}}]{Navid_PRL14}%
  \BibitemOpen
  \bibfield  {author} {\bibinfo {author} {\bibfnamefont {N.}~\bibnamefont
  {Vafaei-Najafabadi}} \emph {et~al.},\ }\bibfield  {title} {\bibinfo {title}
  {Beam loading by distributed injection of electrons in a plasma wakefield
  accelerator},\ }\href {https://doi.org/10.1103/PhysRevLett.112.025001}
  {\bibfield  {journal} {\bibinfo  {journal} {Phys. Rev. Lett.}\ }\textbf
  {\bibinfo {volume} {112}},\ \bibinfo {pages} {025001} (\bibinfo {year}
  {2014})}\BibitemShut {NoStop}%
\bibitem [{\citenamefont {Stupakov}(2013)}]{Stupakov}%
  \BibitemOpen
  \bibfield  {author} {\bibinfo {author} {\bibfnamefont {G.}~\bibnamefont
  {Stupakov}},\ }\bibfield  {title} {\bibinfo {title} {Melting thin foils by
  incident relativistic electron bunch},\ }\href@noop {} {\bibfield  {journal}
  {\bibinfo  {journal} {SLAC-PUB-15729}\ } (\bibinfo {year}
  {2013})}\BibitemShut {NoStop}%
\bibitem [{\citenamefont {Adli}\ \emph {et~al.}(2015)\citenamefont {Adli},
  \citenamefont {Gessner}, \citenamefont {Corde}, \citenamefont {Hogan},\ and\
  \citenamefont {Bjerke}}]{Adli_NIMA15}%
  \BibitemOpen
  \bibfield  {author} {\bibinfo {author} {\bibfnamefont {E.}~\bibnamefont
  {Adli}}, \bibinfo {author} {\bibfnamefont {S.}~\bibnamefont {Gessner}},
  \bibinfo {author} {\bibfnamefont {S.}~\bibnamefont {Corde}}, \bibinfo
  {author} {\bibfnamefont {M.}~\bibnamefont {Hogan}},\ and\ \bibinfo {author}
  {\bibfnamefont {H.}~\bibnamefont {Bjerke}},\ }\bibfield  {title} {\bibinfo
  {title} {Cherenkov light-based beam profiling for ultrarelativistic electron
  beams},\ }\href {https://doi.org/https://doi.org/10.1016/j.nima.2015.02.003}
  {\bibfield  {journal} {\bibinfo  {journal} {Nuclear Instruments and Methods
  in Physics Research Section A: Accelerators, Spectrometers, Detectors and
  Associated Equipment}\ }\textbf {\bibinfo {volume} {783}},\ \bibinfo {pages}
  {35} (\bibinfo {year} {2015})}\BibitemShut {NoStop}%
\bibitem [{\citenamefont {Green}\ \emph {et~al.}(2017)\citenamefont {Green},
  \citenamefont {Hogan}, \citenamefont {Lipkowitz}, \citenamefont {O’Shea},
  \citenamefont {White}, \citenamefont {Yakimenko},\ and\ \citenamefont
  {Yocky}}]{Green_IBIC17}%
  \BibitemOpen
  \bibfield  {author} {\bibinfo {author} {\bibfnamefont {S.~Z.}\ \bibnamefont
  {Green}}, \bibinfo {author} {\bibfnamefont {M.~J.}\ \bibnamefont {Hogan}},
  \bibinfo {author} {\bibfnamefont {N.}~\bibnamefont {Lipkowitz}}, \bibinfo
  {author} {\bibfnamefont {B.}~\bibnamefont {O’Shea}}, \bibinfo {author}
  {\bibfnamefont {G.}~\bibnamefont {White}}, \bibinfo {author} {\bibfnamefont
  {V.}~\bibnamefont {Yakimenko}},\ and\ \bibinfo {author} {\bibfnamefont
  {G.}~\bibnamefont {Yocky}},\ }\bibfield  {title} {\bibinfo {title} {{Beam
  Diagnostic Challenges for FACET-II}},\ }in\ \href
  {https://doi.org/10.18429/JACoW-IBIC2017-MO3AB3} {\emph {\bibinfo {booktitle}
  {Proceedings of the 6th International Beam Instrumentation Conference}}}\
  (\bibinfo {year} {2017})\BibitemShut {NoStop}%
\bibitem [{\citenamefont {Weingartner}\ \emph {et~al.}(2012)\citenamefont
  {Weingartner} \emph {et~al.}}]{Wiengartner_PRSTAB12}%
  \BibitemOpen
  \bibfield  {author} {\bibinfo {author} {\bibfnamefont {R.}~\bibnamefont
  {Weingartner}} \emph {et~al.},\ }\bibfield  {title} {\bibinfo {title}
  {Ultralow emittance electron beams from a laser-wakefield accelerator},\
  }\href {https://doi.org/10.1103/PhysRevSTAB.15.111302} {\bibfield  {journal}
  {\bibinfo  {journal} {Phys. Rev. ST Accel. Beams}\ }\textbf {\bibinfo
  {volume} {15}},\ \bibinfo {pages} {111302} (\bibinfo {year}
  {2012})}\BibitemShut {NoStop}%
\bibitem [{\citenamefont {Barber}\ \emph {et~al.}(2020)\citenamefont {Barber}
  \emph {et~al.}}]{Barber_APL20}%
  \BibitemOpen
  \bibfield  {author} {\bibinfo {author} {\bibfnamefont {S.~K.}\ \bibnamefont
  {Barber}} \emph {et~al.},\ }\bibfield  {title} {\bibinfo {title} {A compact,
  high resolution energy, and emittance diagnostic for electron beams using
  active plasma lenses},\ }\href {https://doi.org/10.1063/5.0005114} {\bibfield
   {journal} {\bibinfo  {journal} {Applied Physics Letters}\ }\textbf {\bibinfo
  {volume} {116}},\ \bibinfo {pages} {234108} (\bibinfo {year}
  {2020})}\BibitemShut {NoStop}%
\bibitem [{\citenamefont {Vafaei-Najafabadi}\ \emph {et~al.}(2016)\citenamefont
  {Vafaei-Najafabadi} \emph {et~al.}}]{Navid_PPCF16}%
  \BibitemOpen
  \bibfield  {author} {\bibinfo {author} {\bibfnamefont {N.}~\bibnamefont
  {Vafaei-Najafabadi}} \emph {et~al.},\ }\bibfield  {title} {\bibinfo {title}
  {Evidence for high-energy and low-emittance electron beams using ionization
  injection of charge in a plasma wakefield accelerator},\ }\href
  {https://doi.org/10.1088/0741-3335/58/3/034009} {\bibfield  {journal}
  {\bibinfo  {journal} {Plasma Physics and Controlled Fusion}\ }\textbf
  {\bibinfo {volume} {58}},\ \bibinfo {pages} {034009} (\bibinfo {year}
  {2016})}\BibitemShut {NoStop}%
\bibitem [{Note1()}]{Note1}%
  \BibitemOpen
  \bibinfo {note} {Manufactured by Mitsubishi Chemical Group}\BibitemShut
  {NoStop}%
\bibitem [{\citenamefont {Dolgashev}\ \emph {et~al.}(2014)\citenamefont
  {Dolgashev} \emph {et~al.}}]{Dolgashev_PRAB14}%
  \BibitemOpen
  \bibfield  {author} {\bibinfo {author} {\bibfnamefont {V.~A.}\ \bibnamefont
  {Dolgashev}} \emph {et~al.},\ }\bibfield  {title} {\bibinfo {title} {Design
  and application of multimegawatt $x$-band deflectors for femtosecond electron
  beam diagnostics},\ }\href {https://doi.org/10.1103/PhysRevSTAB.17.102801}
  {\bibfield  {journal} {\bibinfo  {journal} {Phys. Rev. ST Accel. Beams}\
  }\textbf {\bibinfo {volume} {17}},\ \bibinfo {pages} {102801} (\bibinfo
  {year} {2014})}\BibitemShut {NoStop}%
\bibitem [{\citenamefont {Emma}\ \emph {et~al.}(2018)\citenamefont {Emma},
  \citenamefont {Edelen}, \citenamefont {Hogan}, \citenamefont {O'Shea},
  \citenamefont {White},\ and\ \citenamefont {Yakimenko}}]{Emma_PRAB18}%
  \BibitemOpen
  \bibfield  {author} {\bibinfo {author} {\bibfnamefont {C.}~\bibnamefont
  {Emma}}, \bibinfo {author} {\bibfnamefont {A.}~\bibnamefont {Edelen}},
  \bibinfo {author} {\bibfnamefont {M.~J.}\ \bibnamefont {Hogan}}, \bibinfo
  {author} {\bibfnamefont {B.}~\bibnamefont {O'Shea}}, \bibinfo {author}
  {\bibfnamefont {G.}~\bibnamefont {White}},\ and\ \bibinfo {author}
  {\bibfnamefont {V.}~\bibnamefont {Yakimenko}},\ }\bibfield  {title} {\bibinfo
  {title} {Machine learning-based longitudinal phase space prediction of
  particle accelerators},\ }\href
  {https://doi.org/10.1103/PhysRevAccelBeams.21.112802} {\bibfield  {journal}
  {\bibinfo  {journal} {Phys. Rev. Accel. Beams}\ }\textbf {\bibinfo {volume}
  {21}},\ \bibinfo {pages} {112802} (\bibinfo {year} {2018})}\BibitemShut
  {NoStop}%
\bibitem [{\citenamefont {Hunt-Stone}\ \emph {et~al.}(2021)\citenamefont
  {Hunt-Stone}, \citenamefont {Ariniello}, \citenamefont {Doss}, \citenamefont
  {Lee},\ and\ \citenamefont {Litos}}]{HuntStone_2021}%
  \BibitemOpen
  \bibfield  {author} {\bibinfo {author} {\bibfnamefont {K.}~\bibnamefont
  {Hunt-Stone}}, \bibinfo {author} {\bibfnamefont {R.}~\bibnamefont
  {Ariniello}}, \bibinfo {author} {\bibfnamefont {C.}~\bibnamefont {Doss}},
  \bibinfo {author} {\bibfnamefont {V.}~\bibnamefont {Lee}},\ and\ \bibinfo
  {author} {\bibfnamefont {M.}~\bibnamefont {Litos}},\ }\bibfield  {title}
  {\bibinfo {title} {Electro-optic sampling beam position monitor for
  relativistic electron beams},\ }\href
  {https://doi.org/https://doi.org/10.1016/j.nima.2021.165210} {\bibfield
  {journal} {\bibinfo  {journal} {Nuclear Instruments and Methods in Physics
  Research Section A: Accelerators, Spectrometers, Detectors and Associated
  Equipment}\ }\textbf {\bibinfo {volume} {999}},\ \bibinfo {pages} {165210}
  (\bibinfo {year} {2021})}\BibitemShut {NoStop}%
\bibitem [{\citenamefont {Li}\ and\ \citenamefont {Hogan}(2011)}]{Li_PAC11}%
  \BibitemOpen
  \bibfield  {author} {\bibinfo {author} {\bibfnamefont {S.}~\bibnamefont
  {Li}}\ and\ \bibinfo {author} {\bibfnamefont {M.~J.}\ \bibnamefont {Hogan}},\
  }\bibfield  {title} {\bibinfo {title} {{Beam Diagnostics for FACET}},\ }in\
  \href {https://www.slac.stanford.edu/pubs/slacpubs/14250/slac-pub-14412.pdf}
  {\emph {\bibinfo {booktitle} {Proceedings of the 2011 Particle Accelerator
  Conference}}}\ (\bibinfo {year} {2011})\BibitemShut {NoStop}%
\bibitem [{\citenamefont {Corde}\ \emph {et~al.}(2013)\citenamefont {Corde},
  \citenamefont {Ta~Phuoc}, \citenamefont {Lambert}, \citenamefont {Fitour},
  \citenamefont {Malka}, \citenamefont {Rousse}, \citenamefont {Beck},\ and\
  \citenamefont {Lefebvre}}]{Corde_RMP13}%
  \BibitemOpen
  \bibfield  {author} {\bibinfo {author} {\bibfnamefont {S.}~\bibnamefont
  {Corde}}, \bibinfo {author} {\bibfnamefont {K.}~\bibnamefont {Ta~Phuoc}},
  \bibinfo {author} {\bibfnamefont {G.}~\bibnamefont {Lambert}}, \bibinfo
  {author} {\bibfnamefont {R.}~\bibnamefont {Fitour}}, \bibinfo {author}
  {\bibfnamefont {V.}~\bibnamefont {Malka}}, \bibinfo {author} {\bibfnamefont
  {A.}~\bibnamefont {Rousse}}, \bibinfo {author} {\bibfnamefont
  {A.}~\bibnamefont {Beck}},\ and\ \bibinfo {author} {\bibfnamefont
  {E.}~\bibnamefont {Lefebvre}},\ }\bibfield  {title} {\bibinfo {title}
  {Femtosecond x rays from laser-plasma accelerators},\ }\href
  {https://doi.org/10.1103/RevModPhys.85.1} {\bibfield  {journal} {\bibinfo
  {journal} {Rev. Mod. Phys.}\ }\textbf {\bibinfo {volume} {85}},\ \bibinfo
  {pages} {1} (\bibinfo {year} {2013})}\BibitemShut {NoStop}%
\bibitem [{\citenamefont {{San Miguel Claveria}}\ \emph
  {et~al.}(2019)\citenamefont {{San Miguel Claveria}} \emph
  {et~al.}}]{SanMiguel_RSTA18}%
  \BibitemOpen
  \bibfield  {author} {\bibinfo {author} {\bibfnamefont {P.}~\bibnamefont {{San
  Miguel Claveria}}} \emph {et~al.},\ }\bibfield  {title} {\bibinfo {title}
  {Betatron radiation and emittance growth in plasma wakefield accelerators},\
  }\href {https://doi.org/10.1098/rsta.2018.0173} {\bibfield  {journal}
  {\bibinfo  {journal} {Philosophical Transactions of the Royal Society A:
  Mathematical, Physical and Engineering Sciences}\ } (\bibinfo {year}
  {2019})}\BibitemShut {NoStop}%
\bibitem [{Note2()}]{Note2}%
  \BibitemOpen
  \bibinfo {note} {Manufactured by Epic-Crystal}\BibitemShut {NoStop}%
\bibitem [{\citenamefont {Ross}(1928)}]{Ross_JOptSocAm28}%
  \BibitemOpen
  \bibfield  {author} {\bibinfo {author} {\bibfnamefont {P.~A.}\ \bibnamefont
  {Ross}},\ }\bibfield  {title} {\bibinfo {title} {A new method of spectroscopy
  for faint x-radiations},\ }\href {https://doi.org/10.1364/JOSA.16.000433}
  {\bibfield  {journal} {\bibinfo  {journal} {J. Opt. Soc. Am.}\ }\textbf
  {\bibinfo {volume} {16}},\ \bibinfo {pages} {433} (\bibinfo {year}
  {1928})}\BibitemShut {NoStop}%
\bibitem [{\citenamefont {Agostinelli}\ \emph {et~al.}(2003)\citenamefont
  {Agostinelli} \emph {et~al.}}]{GEANT}%
  \BibitemOpen
  \bibfield  {author} {\bibinfo {author} {\bibfnamefont {S.}~\bibnamefont
  {Agostinelli}} \emph {et~al.},\ }\bibfield  {title} {\bibinfo {title}
  {Geant4—a simulation toolkit},\ }\href
  {https://doi.org/https://doi.org/10.1016/S0168-9002(03)01368-8} {\bibfield
  {journal} {\bibinfo  {journal} {Nuclear Instruments and Methods in Physics
  Research Section A: Accelerators, Spectrometers, Detectors and Associated
  Equipment}\ }\textbf {\bibinfo {volume} {506}},\ \bibinfo {pages} {250}
  (\bibinfo {year} {2003})}\BibitemShut {NoStop}%
\bibitem [{\citenamefont {{San Miguel Claveria}}\ \emph
  {et~al.}(2022)\citenamefont {{San Miguel Claveria}} \emph
  {et~al.}}]{SanMiguel_AAC}%
  \BibitemOpen
  \bibfield  {author} {\bibinfo {author} {\bibfnamefont {P.}~\bibnamefont {{San
  Miguel Claveria}}} \emph {et~al.},\ }\bibfield  {title} {\bibinfo {title}
  {{Commissioning and first measurements of the initial X-ray and $\gamma$-ray
  detectors at FACET-II}},\ }in\ \href@noop {} {\emph {\bibinfo {booktitle}
  {Proceedings of the 20th Advanced Accelerator Concepts Workshop
  (AAC’22)}}}\ (\bibinfo {year} {2022})\BibitemShut {NoStop}%
\bibitem [{\citenamefont {Naranjo}\ \emph {et~al.}(2021)\citenamefont {Naranjo}
  \emph {et~al.}}]{Naranjo_IPAC21}%
  \BibitemOpen
  \bibfield  {author} {\bibinfo {author} {\bibfnamefont {B.}~\bibnamefont
  {Naranjo}} \emph {et~al.},\ }\bibfield  {title} {\bibinfo {title} {{Compton
  Spectrometer for FACET-II}},\ }in\ \href
  {https://doi.org/10.18429/JACoW-IPAC2021-THPAB269} {\emph {\bibinfo
  {booktitle} {Proceedings of the 12th International Particle Accelerator
  Conference}}}\ (\bibinfo {year} {2021})\BibitemShut {NoStop}%
\bibitem [{\citenamefont {Joshi}(2006)}]{Joshi_SA06}%
  \BibitemOpen
  \bibfield  {author} {\bibinfo {author} {\bibfnamefont {C.}~\bibnamefont
  {Joshi}},\ }\bibfield  {title} {\bibinfo {title} {{Plasma Accelerators}},\
  }\href {http://www.jstor.org/stable/26061335} {\bibfield  {journal} {\bibinfo
   {journal} {Scientific American}\ }\textbf {\bibinfo {volume} {294}},\
  \bibinfo {pages} {40} (\bibinfo {year} {2006})}\BibitemShut {NoStop}%
\bibitem [{\citenamefont {Zhang}\ \emph {et~al.}(2023)\citenamefont {Zhang},
  \citenamefont {Storey} \emph {et~al.}}]{ChaojiePaper}%
  \BibitemOpen
  \bibfield  {author} {\bibinfo {author} {\bibfnamefont {C.~J.}\ \bibnamefont
  {Zhang}}, \bibinfo {author} {\bibfnamefont {D.}~\bibnamefont {Storey}}, \emph
  {et~al.},\ }\bibfield  {title} {\bibinfo {title} {{Generation of meter-scale
  hydrogen plasmas and efficient pump limited wakeﬁeld excitation using 10
  GeV electron bunches}}} (\bibinfo {year} {2023}),\ \bibinfo {note} {submitted
  for publication}\BibitemShut {NoStop}%
\end{thebibliography}%
